\documentstyle[12pt,subeqn]{article}
\newcommand{\beeq}{\begin{equation}}
\newcommand{\eneq}{\end{equation}}
\newcommand{\beqn}{\begin{eqnarray}}
\newcommand{\eeqn}{\end{eqnarray}}
\def\la{\langle}
\def\ra{\rangle}
\def\psibar{\bar{\psi}}
\def\psid{\psi ^{\dagger}}
\def\kna{N,{\bar A}\rangle}

\def\knap1{N'_1,\bar{A}'_1\rangle}
\def\bnap1{\langle N'_1,\bar{A}'_1}
\def\kn1a{N_1,\bar{A}_1\rangle}
\def\bn1a{\langle N_1,\bar{A}_1}
\def\kE{E_n\rangle}
\def\bE{\langle E_n}
\def\kEp{E_{n'}\rangle}
\def\bEp{\langle E_{n'}}
\def\tx{\tilde{x}}
\def\ty{\tilde{y}}

\def\td{\tilde{\delta}}
\def\tg{\tilde{\gamma}}
\def\next{{~~~,~~~}}
\def\ad{a^{\dag}}
\def\onehalf{\hbox{${1\over 2}$}}
\def\pardx{\partial_{\bar{x}}}
\def\pardt{\partial_{\bar{t}}}
\def\gbm{\overline{G}_{\beta,m}(\bar{x},\bar{t})}
\def\tgbm{\tilde{G}_{\beta,m}(\bar{x},\bar{t})}
\def\ngbm{\overline{G}_{\beta,m}^{(0)}(\bar{t})}
\def\rab{\rangle_{\beta}}
\def\rabv{\rangle_{\beta, vac}}
\def\rabe{\rangle_{\beta, exc}}

\begin{document}

\title{The Schwinger Model on a Circle: Relation between Path Integral and
Hamiltonian approaches}

\author{S.Azakov \footnote{e-mail address: azakov\_s@hotmail.com}}
\maketitle

{\it Institute of Physics, Azerbaijan National Academy of Sciences, 
 Baku,
Azerbaijan}
\begin{center}
{\it and}\\
\end{center}
{\it The Abdus Salam International Center for Theoretical Physics, 
Trieste,
Italy}\\

\begin{abstract}
We solve the massless Schwinger model exactly in Hamiltonian formalism on
a circle. We construct physical states explicitly and discuss the role of
the spectral flow and nonperturbative vacua. Different thermodynamical
correlation functions are calculated and after performing the analytical
continuation are compared with the corresponding expressions obtained for
the Schwinger model on the torus in Euclidean Path Integral formalism 
obtained before.
\end{abstract}

\section{Introduction}
It is well-known that systems composed by Dirac fields and gauge fields
possess a very intricate `non-perturbative structure'(NPS). An essential
part of this NPS is determined by topological properties of gauge field
configurations and spinor field configurations, and their relations described
by `Index theorems'.  It is the aim of this paper to describe the
non-perturbative structure for a simple example as precisely and transparently
as possible. This model is $U(1)$ gauge theory with massless fermions in two
space-time dimensions, the so-called Schwinger model (SM) \cite {SM}. We consider this
model in the Hamiltonian formulation in $1+1$  dim. Minkowski space-time and compare obtained
results with the corresponding results obtained before
in the path integral formulation in Euclidean space. Our central point will
be the description of the relation of these two approaches. There are many
occasions where a better understanding of this relation is highly desirable.
We are convinced that the necessary preciseness in the description of the
NPS can be achieved only if it is considered in a limit of field theories on
compact spaces \cite {Joos}.
We choose as starting points the Hamiltonian formulation of
the SM on a circle $S_1 = \{y| 0 \leq y < L \}$,
and its path integral formulation on the torus ${\cal T}_2 
= \{x_{\mu}| 0 \leq x_{\mu} \leq L_{\mu}; \mu = 1,2 \}$.
It is important that the result of both treatments are explicitely documented
in literature. For the
discussion of their relation, one has to show that the
thermodynamical expectation values are related to the corresponding expressions
in path integral approach by analytical continuation.
In ordinary field theory, this is generally treated by methods related to the
Osterwalder-Schrader Theorem. Here we want to discuss this topic for the
SM including the topological features of NPS on compact spaces.
In more complex, more physical gauge theories like QCD, such questions
are treated in a less transparent manner under the heading of the
`vacuum tunneling picture'.

The following Table should give an overview of our progam. It should be
understandable for people familiar with the literature. Otherwise we will
quote the relevant results below.
\newpage
\begin{tabular}{|l|l|}       \hline \hline
Euclidean Path Integral &  Hamiltonian Formulation  \\ \hline \hline
General path integral formula & Canonical formalism and \\
& thermodinamical expectation values \\ \hline
Topological quantum number & Non-trivial gauge family                 \\ \hline
Atiyah-Singer Theorem &    Spectral flow theorem     \\  \hline
Effective action for the gauge field & Dirac sea construction \\
                             & Current algebra        \\
                             & Bosonized fermion operator \\ \hline
Regularization, point splitting & Regularized charges and Hamiltonian \\
\hline
 Invariant gauge field integration &
Quantum theory of \\
& gauge field in temporal gauge \\
  & Manton condition       \\
& Gauss constraint \\
& Bogoliubov transformation \\ \hline
Z factor   & Partition function \\  \hline
Gauge field propagator & Thermodynamical expectation values \\
& of the $T$ products of scalar fields  \\  \hline
Instanton zero mode contribution &  Excited nonperturbative \\
&vacuum \\ \hline
Fermionic condensate & Thermodynamical expectation values  \\
 $ \la \bar{\psi}(x)\psi(x) \ra \;\;\;
 \rightarrow $ &
$ \leftarrow \; \la \bar{\psi}(x)\psi(x) \ra_{\beta}$ \\
Currents correlation function&   \\
 $  \la j_{\mu}(x)j_{\nu}(x') \ra  \;\;\; \rightarrow $
 &$\leftarrow \;   \la T j_{\mu}(x)j_{\nu}(x') \ra_{\beta} $ \\
Densities  correlation function & \\
 $\la \bar{\psi}(x)\psi(x)\bar{\psi}(x')\psi(x') \ra \;\;
\rightarrow
$ & $\leftarrow \;
\la T\bar{\psi}(x)\psi(x)\bar{\psi}(x')\psi(x') \ra_{\beta} $
\\  \hline \hline
\end{tabular}
\\

The paper is organized as follows. In Section 2 we briefly review the
results obtained before for the SM on a circle in Hamiltonian approach \cite {Man},\cite{IM},\cite{HH}
and relevant for the present consideration.
In addition to these results we give some new information
which concerns gauge invariant states and the expressions for observables
functions in the nontrivial topological sectors. Section 3 is devoted to the
canonical calculations of thermodinamical expectation values and important correlation functions.
In section 4 we show how expressions obtained in Hamiltonian and Path integral approach \cite{SW},
\cite{Azakov},\cite{JA},\cite{Azakov2} relate to each other. Some technical details are given in
the Appendices.

\section{Canonical treatment of the SM on a circle}

In this section we want to give a compilation of quantum mechanical
ingredience of the SM on a circle.
For further details we refer to the quoted literature
\cite{Man},\cite{IM},\cite{HH}.
The starting point is the Hamiltonian in temporal gauge
($A_0(y)=0,~ A_1(y)\equiv A(y), \alpha = \gamma_{5}$ and $ F(y)$ is an
electric field ):
\begin{eqnarray}
H & = &\int_0^L dy \left\{\frac{F^2(y)}{2} + \frac{1}{i} \psi^{\dagger}(y)
\alpha (\partial_y -ie A(y))\psi(y)\right\} \nonumber  \\
&=& H_F + H_q
\label{hcan}
\end{eqnarray}
together with the canonical CR
\begin{displaymath}
[F(y),A(y^{\prime})]_{-}  =
\frac{1}{i} \delta(y\,-\, y^{\prime}),\;\;
[F(y),F(y^{\prime})]_- =\ldots=0,
\end{displaymath}
\begin{displaymath}
\{\psi_{\alpha}(y),\psi_{\beta}^{\dagger}(y^{\prime})\}
= \delta(y\!-\! y^{\prime})\delta_{\alpha \beta},~~~~
\{\psi_{\alpha}(y),\psi_{\beta}(y^{\prime})\} = \ldots = 0,
\end{displaymath}
and the Gauss law as a constraint  $\partial_yF(y)
+e\psi^{\dagger}(y) \psi(y)=0$. $L$ is a circumference of a circle
$S_1$ and $\delta (y)$ is Dirac's $\delta$ -function on it.

The Gauss law is implemented by the gauge transformations generated by
\begin{equation}
G[\lambda_0] = -\frac{1}{e}\int_0^L dy\,(\partial_y\,F(y)
+ e\psi^{\dagger}(y)\psi(y))\lambda_0(y).\label{gauss}
\end{equation}
The infinitesimal gauge transformation for $\lambda_0(0) = \lambda_0(L)$
 follow from the CR
\begin{eqnarray}
i[G[\lambda_0],A(y)] \!\! &=& \!\!
\frac{1}{e}\partial_y\lambda_0(y),\;\;\;\;\;\;
 i[G[\lambda_0],F(y)] = 0, \label{gauge} \\
i[G[\lambda_0],\psi(y)] &\!\!=&\!\! i\lambda_0(y)\, \psi(y),\;\;\;
i[G[\lambda_0],\psi^{\dagger}(y)]
= - i\lambda_0(y)\, \psi^{\dagger}(y). \nonumber
\end{eqnarray}
 The topology of $S_1$ induces a classification of gauge
transformations $\Lambda[\lambda_n(y)] $ according to winding numbers $n$:
 $ \Lambda[\lambda_n(y)] = e^{i\lambda_n(y)}$ with $\lambda_n(L)
-\lambda_n(0)
=2\pi n ,\;\; n $ integer. If $n \neq 0\; (n=0)$
we call $\Lambda[\lambda_n(y)]$ a large (small) gauge transformation.
A general {\bf large gauge transformation} is a product of a special
large gauge transformation, i.e. $\Lambda_n =e^{2\pi i n y/L} $, with a
small gauge transformation. In particular $\Lambda_n$ transforms a
constant gauge field $A=\int_0^LA(y)dy$:
\begin{equation}
\Lambda_n (A) = A + \frac{2\pi n}{e} ~.
\end{equation}
Therefore $ eA $ in the interval $0 \leq eA < 2\pi $ is a gauge invariant
 quantity. It represents
a topological non-trivial family in the space of gauge invariants.
\subsection{Spectral flow and vacuum structure}
In the following we consider first $H_q$ as the Hamiltonian of fermions
in an external gauge field. For this we regard the expansion of
 $\psi(y) ,\bar{\psi}(y)$ in normal modes of the `single particle
Hamiltonian ${\cal H}$':
\begin{equation}\label{ham1}
{\cal H}\phi_k(y)  =
 \frac{1}{i} \alpha(\partial_y -ieA(y))\phi_k(y)
 = E_k\phi_k(y).
\end{equation}
For periodic boundary conditions the solutions of the eigenvalue equation
are
\begin{eqnarray}
\phi_{R,k}(y) &=& \phi_k(y)
\left(\begin{array}{c} 1\\0 \end{array}\right), \;\;{\rm with}\;\;
E_{R,k}= E_k \equiv \frac{2\pi}{L}(k - \bar{A}),\nonumber \\
\phi_{L,k}(y) &=& \phi_k(y)
\left(\begin{array}{c} 0\\1 \end{array}\right), \;\;{\rm with}\;\;
E_{L,k}= -E_k,\nonumber \\
\phi_k(y)&=& L^{-1/2} \exp \left(2\pi i\frac{y}{L}k + ie\int_0^y A(y')dy'
-2\pi i \bar{A}\frac{y}{L} \right),\nonumber \\
\bar{A} &\equiv&\frac{e}{2\pi}\int_0^L A(y)dy ,\;\; k \;
{\rm integer} \label{eigf}.
\end{eqnarray}
The spectrum of ${\cal H}$ shows the phenomenon of
{\bf `spectral flow'}:
 When $\bar{A}$
varies between the gauge equivalent values: $ 0\rightarrow 1$,
  then the positive chirality energy decreases:
$E_{R,k} \rightarrow E_{R,k}-2\pi/L $ , and the negative chirality energy
increases: $ E_{L,k} \rightarrow E_{L,k}+2\pi/L $ \cite{Man}.

With help of these wave functions the fermion operators are represented
by creation and annihilation (CA) operators
\begin{eqnarray}
\psi(y)= \psi_R(y) + \psi_L(y) &= & \sum_{k} \left( a(k)\phi_{R,k}(y)+
b(k)\phi_{L,k}(y) \right) ,
\nonumber \\
\psi^{\dagger}(y)= \psi^{\dagger}_R(y)+\psi^{\dagger}_L(y)&= & \sum_{k}
\left(
a^{\dagger}(k)\phi^{*}_{R,k}(y)
 + b^{\dagger}(k)
\phi^{*}_{L,k}(y) \right).
\end{eqnarray}
Spectral flow leads to an involved `Dirac sea construction' of the
vacuum
state which determines the representation of the CR:
\begin{displaymath}
\{a(k),a^{\dagger}(k')\}  = \{b(k),b^{\dagger}(k')\} =
\delta_{k,k'}, \end{displaymath}
 \begin{equation}
\{a(k),a(k')\}  =
\{b(k),b(k')\} = \dots = 0. \label{CR}
\end{equation}
We define a `relative Dirac sea state' (RDSS) :  $|N_+,N_-;\bar{A}\ra $
in which all energy
levels $E_{R,k}<E_{R,N_+} (E_{L,k}\leq E_{L,N_-}) $ with chirality
$+(-)$ are occupied, and with the other levels empty:
\begin{eqnarray} \label{vacst}
 a^{\dagger}(k)|N_+,N_-;\bar{A}\ra &=& 0 \;\;{\rm for} \;\; k\leq N_+ -1,
 \nonumber \\
a(k) |N_+,N_-;\bar{A}\ra & = &0 \;\;{\rm for} \;\; k\geq N_+ ,\nonumber \\
b(k) |N_+,N_-;\bar{A}\ra &=& 0 \;\;{\rm for} \;\; k\leq N_- -1,
\nonumber\\
b^{\dagger}(k) |N_+,N_-;\bar{A} \ra& = & 0 \;\;{\rm for} \;\; k\geq N_-.
\end{eqnarray}
 Gauss law implies that on $S_1$ the {\it total charge is zero}. This
means $N_+ = N_- $ as we will show below.
\par
We introduce the Fourier decomposition for the currents
with positive and negative chirality
\begin{equation}
j_{\pm}(y) = \frac{1}{2}\psi^{\dagger}(y)(1\pm\alpha)\psi(y)
= \frac{Q_{\pm}}{L} + \frac{1}{L}\sum_{k\not=0}j_{\pm}(k)
e^{2\pi i k \frac{y}{L}}.
\label{currx}
\end{equation}
In terms of $a$ and $b$ operators $j_{\pm}(k)$ can be written as follows:
\begin{eqnarray}
j_{+}(k)= \sum_n a^{\dagger}(n) a(n+k), \nonumber \\
j_{-}(k)= \sum_n b^{\dagger}(n) b(n+k).
\label{currk+-}
\end{eqnarray}
A careful calculation shows \cite{IM},
that on the RDSS they satisfy the CR of current algebra
\begin{equation}
\label{cra}
[j_{\pm}(k),j^{\dagger}_{\pm}(k')] = \pm k\delta_{k,k'},\;\;
k,k' > 0, \;\;\;j_{\pm}(k)=j^{\dagger}_{\pm}(-k).
\end{equation}
\subsection{Regularized charges and regularized Hamiltonian}
The chiral charge operators $Q_{\pm}$,
as well as the fermion part of the
Hamilton operator $H_q$ might be expressed
by the CA operators. In order to make them well defined they must be
`Wick ordered' with respect to the RDSS $|N_+,N_-;\bar{A} \rangle $. For
example
\begin{eqnarray}
Q_+ & = & \frac{1}{2}\int_0^Ldy \psi^{\dagger}(y)(1+\alpha)
\psi(y)\nonumber \\
&=& \sum_k a^{\dagger}(k)a(k)
 = \sum_{k= N_+}^{\infty}a^{\dagger}(k)a(k)
\nonumber \\ & &
-\sum_{k=-\infty}^{N_+ -1}a(k)a^{\dagger}(k)
+ \sum_{k=-\infty}^{N_+ -1}1\nonumber \\
&\equiv & {\cal N }_+[\sum_k a^{\dagger}(k)a(k)] +
\lim_{s\rightarrow 0} \sum_{k= - \infty}^{N_+ -1}
|\lambda E_k|^{-s}.
\end{eqnarray}
The constant, i.e. the expectation values $\langle Q_{\pm}\rangle $
 in the RDSS, must be regularized.
We choose the $\zeta$-function regularization, as we have indicated in the last
line. The result is
\begin{eqnarray}
\la Q \ra & =& \la Q_+ +  Q_-\ra = N_+ - N_- = 0,
 \nonumber \\
 \la Q_5 \ra& = &\la Q_+ - Q_- \ra
 =N_+ + N_- -1 - 2\bar{A}.
\end{eqnarray}
 In the following we restrict ourself to `electro-magnetically
 neutral' RDSS: $ \la Q \ra  = 0 $, i.e. $ N_+=N_- = N $.

Similarly we get for the fermionic Hamiltonian
\begin{eqnarray}\label{ham5}
H_q  &=& \sum_{k}\{ E_ka^{\dagger}(k)a(k)
 - E_kb^{\dagger}(k)b(k)\} \nonumber \\
& =& \sum_{k=N_+}^{\infty} E_ka^{\dagger}(k)a(k)
-\sum_{k=-\infty}^{N_+ -1}E_ka(k)a^{\dagger}(k)
 +\lim_{s\rightarrow 0}\sum_{k=-\infty}^{N_+-1}E_k|E_k|^{-s} \nonumber \\
&-&\sum_{k=-\infty}^{N_--1}E_kb^{\dagger}(k)b(k)
+\sum_{k= N_-}^{\infty} E_kb(k)b^{\dagger}(k)
- \lim_{s\rightarrow 0}  \sum_{k=N_-}^{\infty}E_k|E_k|^{-s}  \nonumber \\
&\equiv  & :H_q:_N + \la E \ra_{{\rm reg}},
\end{eqnarray}
with  $ \la E \ra_{{\rm reg}} =\frac{2\pi}{L}\left[
 (\bar{A} - N +\frac{1}{2})^2  - \frac{1}{12}\right]. $
It is the dependence of $Q_5$ on $\bar{A}$
which leads to the Heisenberg equation: $ \dot{Q}_5 = i[H,Q_5]
= - \frac{e}{\pi}LF$, where $F$ is the constant part of the
electric field $F(y)$.\\
{\bf This implies that the chiral charge is not
conserved (`chiral anomaly')}. It has its origin in the phenomenon
of spectral flow.
\par
It turns out \cite{Man}, \cite{IM} that on the space generated by
$j_{\pm}(k)\;\; $,
applied to the RDSS the following expression describes the same
excitations
as $:H_q:_N$
\begin{equation}\label{ham3}
:H_q:_N = \frac{2\pi}{L}\sum_{k>0}
(j_+^{\dagger}(k)j_+(k) +
j_-(k)j_-^{\dagger}(k))~.
\end{equation}
Thus on this space we identify $:H_q:_N$ with this Sugawara form
~(\ref{ham3}).

\subsection{Diagonalization of the total Hamiltonian}
For the diagonalization of  the total Hamiltonian on the sub-space
of gauge invariant states
 we have to include the effect of the
 part $H_F$ of the Hamiltonian Eq.(\ref{hcan}) depending on the gauge fields.
 First we introduce the Fourier
 decomposition of the gauge fields
\beeq
F(y)= F + \frac{1}{L}\sum_{k \not= 0}f(k)e^{2\pi i k \frac{y}{L}},\;
 \;\;\; f^{\dagger}(k) = f(-k),
\label{elecf}
\eneq
\beeq
A(y)= \frac{2\pi}{eL}\bar{A} +
\frac{1}{L}\sum_{k \not= 0}A(k)e^{2\pi i k \frac{y}{L}},\;
 \;\;\;A^*(k) = A(-k).
\label{vecpot}
\eneq
On gauge invariant states we may use the Gauss condition which reads in
Fourier components
\begin{equation}
\frac{2\pi k i}{L}f(k) +e(j_+(k) +j_-(k)) =  0. \label{GL}
\end{equation}
It allows for $k \neq 0 $ the elimination of the Fourier component of
$F(y)$ in $H_F$, and leads to the introduction of the Coulomb energy:
\begin{eqnarray}
H_F& = &\frac{1}{2}\int_0^LF^2(y)dy  = \frac{L}{2}F^2 +
\frac{1}{L}\sum_{k>0}f^{\dagger}(k)f(k) \nonumber  \\
& = & \frac{L}{2}F^2 + \frac{e^2 L^2}{4\pi^2}\sum_{k > 0 }
\frac{1}{k^2} \left(j_+^{\dagger}(k)+ j_-^{\dagger}(k)\right)\left(j_+(k)+
j_-(k)\right).
\end{eqnarray}
Adding to $H_F$ the fermionic Hamiltonian with vacuum part and Sugawara form
of the excitation, we may write $H$, Eq.(\ref{hcan}) , as
\begin{eqnarray}
H& = & \frac{L}{2} F^2 + \frac{2\pi}{L}\left[
 \left(\bar{A} - N +\frac{1}{2}\right)^2 -\frac{1}{12}\right] \nonumber \\
& & + \frac{1}{L} \sum_{k>0}
 \{j_a^{\dagger}(k){\cal M}_{ab}(k) j_b(k)- 2\pi k \} \nonumber \\
& = & H_{vac} + H_{exc}, \label{hamil}
\end{eqnarray}
with
$ {\cal M}_{+-} = {\cal M}_{-+}=e^2L^2/4 \pi^2 k^2 \;\;
{\cal M}_{++}= {\cal M}_{--} =2\pi + e^2L^2/4 \pi^2 k^2$.

Let us first treat $H_{vac}$. It acts on the RDSS of Eq.~(\ref{vacst})
with the constant potential $\bar{A}$, as a parameter.
In quantum mechanical language, we consider $\bar{A}$
diagonal: operator $\bar{A}$ acting on $|N,\bar{A}\ra$ gives c-number
$\bar{A}$ miltiplied $|N,\bar{A}\ra$.\\
With the ansatz Eqs.(\ref{ham1}),(\ref{eigf}) we have gauged away space
dependent
 components of $A(y)$.
However, the phenomenon of spectral flow makes these states not invariant
under large gauge transformation
\begin{equation}
U^{-1}|N,\bar{A}\rangle = |N+1,\bar{A}+1\rangle.
\end{equation}
According to Manton, gauge invariant states must be a superposition
\begin{equation}
|\omega \ra = \int_0^1 d \{\bar{A}\} \sum_N
\Psi_N(\{\bar{A}\})|N,\{\bar{A}\}\ra,
\label{vgis}
\end{equation}
where $\{\bar{A}\}$ is a fractional part of the electromagnetic
potential's global part which we considered before :$\bar{A} =
[\bar{A}] + \{\bar{A} \}$, and  $[\bar{A}]$ is an integer part.
Note that $\{\bar{A}\} \in [0,1)$ is invariant under large gauge
transformations , so it is invariant under all gauge transformations.
We will continue the function $\Psi_N(\{\bar{A}\})$ to the whole interval
$[0,1]$ using so called Manton's `periodicity conditions ':
\begin{equation}
\Psi_{N+1}(1) = \Psi_N(0), \;\;\;\;
\Psi_{N+1}'(1) = \Psi_N'(0),
\label{pc}
\end{equation}
(prime means derivative)
because the states $|N,\bar{A}=0 \ra$ and $|N+1, \bar{A}=1\ra$ are gauge
equivalent, and the spectrum flow makes the transition smooth.\\
According to the Hamiltonian Eq.(\ref{hamil}), the wave function
$\Psi_N(\bar{A})$ which describes its eigenstate for fixed $N$ and $0\le
\bar{A} \le 1$ must satisfy the
Schroedinger equation
\begin{equation}
\left\{ -\frac{e^2L}{8\pi^2} \frac{d^2}{d\bar{A}^2} + \frac{2\pi}{L}\left[
 \left(\bar{A} - N + 1/2\right)^2
-\frac{1}{12}\right]\right\}\Psi_N(\bar{A})
= E\Psi_N(\bar{A}).
\label{SE}
\end{equation}
It is of the oscillator type. Its normalized solutions are
 \begin{equation}
\Psi_{N,n}(\bar{A}) = \left(\frac{\omega}{\pi}\right)^{1/4}
\frac{1}{(2^n n!)^{1/2}} H_n(\sqrt{\omega}(\bar{A}-N + 1/2))
e^{-\frac{\omega}{2}(\bar{A}-N+ 1/2)^2},
\label{Psi}
\end{equation}
with $ \omega \equiv 4 \pi^{3/2}/eL \equiv 4\pi/m L $, $H_n$ denotes the Hermite
polynomial,
and the energy eigenvalues
\begin{equation}
E_n = m(n+ 1/2) - \frac{\pi}{6L}. \label{vspec}
\end{equation}
The wave functions (\ref{Psi}) obey Manton's periodicity conditions
(\ref{pc}).
The physical gauge invariant ground state of $H_{vac}$which we call
physical vacuum  is therefore
\begin{eqnarray}
|{\rm phys.vac.} \ra =\sum_N\int_0^1d\bar{A}
\left(\frac{\omega}{\pi}\right)^{1/4}
e^{-\frac{\omega}{2}(\bar{A}-N+1/2)^2}|N,\bar{A}\ra.
\end{eqnarray}
Now we will treat $H_{exc}$.  We can diagonalize
$j_a^{\dagger}(k){\cal M}_{ab}(k) j_b(k)$ with the help of a Bogoliubov
transformation
\begin{displaymath}
A(k) = \frac{1}{\sqrt{k}}( j_+(k) \cosh \alpha(k)
+j_-(k)\sinh \alpha (k)) , \end{displaymath}
\begin{equation}
B^{\dagger}(k) = \frac{1}{\sqrt{k}}(j_+(k) \sinh \alpha(k)
+j_-(k) \cosh \alpha (k)),
\label{bt}
\end{equation}
with
\begin{eqnarray}
\cosh 2\alpha(k)& = &\frac{1}{E(k)}(2\pi k/L + e^2L/(4 \pi^2 k)),
\nonumber\\
\sinh 2\alpha(k)& = & e^2L/(4 \pi^2 E(k)k),\nonumber\\
E(k)& =&\sqrt{\left(\frac{2\pi k}{L}\right)^2 +\frac{e^2}{\pi}}.
\label{alpha}
\end{eqnarray}
 These operators $A(k), B(k)$ etc. satisfy the usual canonical CR:
\[ [A(k),A^{\dagger}(k')] = \delta_{kk'},\ldots \]
The Bogoliubov transformation can be implemented in the usual manner by
the unitary operator
: ${\cal U} = \prod_k U(k) $, with
\[ U(k)(  j_+(k), j_-(k))U^{-1}(k) = \sqrt{k}(A(k),B^{\dagger}(k))~,\]
\begin{equation}
U(k)=
\exp\left\{-\frac{\alpha(k)}{k}\left
(j^{\dagger}_+(k)j_-(k)-j_+(k)j^{\dagger}_-(k)\right )
\right\}.
\end{equation}
 Adding up the partial results of this Section, we get for the transformed
total Hamiltonian, Eq.(\ref{hcan})
\begin{eqnarray} \label{feq}
{\cal U}H{\cal U}^{-1}& = & \frac{L}{2} F^2 + \frac{\pi}{L}\left [
\frac{ Q_5^2}{2}
 -\frac{1}{6}\right] \nonumber \\
& + & \sum_{k> 0}\left \{E(k)\left(A^{\dagger}(k)A(k) +
B^{\dagger}(k)B(k)\right)
 + E(k) -\frac{2\pi k}{L}\right \} \nonumber \\
&=& H_{vac}+{\tilde H}_{exc} ~~. 
\label{hamil2}
\end{eqnarray}
This expression contains the main result of the canonical treatment of the
SM. The first term describes an intricate
vacuum structure as discussed above. The second term describes right
 and left moving  massive free particles on the circle.

\subsection{The algebra of observables}
Of course we should make some
remarks on how fermions are described
by observables. There is a general scheme for extending the
physical
Hilbert space generated by local observables to a larger
space in which
field operators relatively local to the observables are
represented \cite{BMT}.
 For currents on a circle satisfying an algebra with CR
like above Eq.(\ref{cra}), we get
\begin{equation}
j_{\pm}(x)= \frac{Q_{\pm}}{L}\mp \frac{1}{\sqrt{\pi}}\partial_x
\tilde{\varphi}_{\pm}(x)~,
\end{equation}
where the scalar fields $\tilde{\varphi}_{\pm}(x)$ are defined in
Eqs.(\ref{phiplus}) and (\ref{phiminus}).

We have the bosonized expressions for the fermion fields (see
Eqs.(\ref{bos1}) and (\ref{O_+1}))
\begin{eqnarray}
\psi_{R}(x)&=&
\frac{1}{\sqrt{L}}C_+U_+e^{2\pi i\frac{x}{L}Q_+ -i\pi\frac{x}{L}
+ie \int_0^x A(x')dx'}e^{-{\cal A}^{\dag}(x)}e^{{\cal A}(x)} \nonumber \\
&=&\frac{1}{\sqrt{L}}C_+U_+ e^{2\pi i\frac{x}{L}Q_+ -i\pi\frac{x}{L}
+ie \int_0^x A(x')dx'}
:e^{-i2\sqrt{\pi}\tilde{\varphi}_{+}(x)}:~,
\label{bosonizR}
\end{eqnarray}
and
\begin{eqnarray}
\psi_{L}(x)&=&
\frac{1}{\sqrt{L}}C_-U_-e^{-2\pi i\frac{x}{L}Q_- +i\pi\frac{x}{L}
+ie \int_0^x A(x')dx'}e^{-{\cal B}^{\dag}(x)}e^{{\cal B}(x)} \nonumber \\
&=&\frac{1}{\sqrt{L}}C_-U_- e^{-2\pi i\frac{x}{L}Q_- +i\pi\frac{x}{L}
+ie \int_0^x A(x')dx'}
:e^{-i2\sqrt{\pi}\tilde{\varphi}_{-}(x)}:~,
\label{bosonizL}
\end{eqnarray}
where the operators $C_{-},U_{\pm}, {\cal A}(x)$ and ${\cal B}(x)$ are
defined in
Appendix A (Eqs.(\ref{KF}), (\ref{operU_+}), (\ref{operU_-}) and
 (\ref{caligA}), (\ref{caligB})) and $C_{+}=1$.

Using bosonization formulae (\ref{bosonizR}),(\ref{bosonizL}) and the
relation between operators
$j_{\pm}(k)$ and operators $A(k)$ and $B(k)$ which are obtained after
Bogoliubov
transformation (for $(k>0)$)
\begin{eqnarray}
j_+(k) = \sqrt{k}[A(k)\cosh\alpha(k) - B^{\dag}(k) \sinh\alpha(k)]~,
\nonumber \\
j_-(k) = \sqrt{k}[B^{\dag}(k)\cosh\alpha(k) - A(k) \sinh\alpha(k)]~,
\label{currk}
\end{eqnarray}
we get for the {\it chiral operator}
\begin{eqnarray} \label{bosonf}
\psi_R^{\dag}(x)\psi_L(x)= -\frac{1}{L}e^{-i\pi(2\frac{x}{L} -1)
(Q_++Q_-)}U_+^{\dag}U_-  \\ \nonumber
 \times \exp\sum_{k>0}\left
\{\frac{1}{k}+\beta_x(k)A^{\dag}(k)-\beta^*_x(k) B^{\dag}(k)
-\beta^*_x(k)A(k)+\beta_x(k)B(k)\right\},
\end{eqnarray}
where
\begin{equation}
\beta_x(k) \equiv \frac{1}{\sqrt{k}}\left[\cosh \alpha(k) - \sinh \alpha(k) \right]
e^{-2\pi ik\frac{x}{L}}~.
\label{betax}
\end{equation}
and for the {\it gauge invariant fermionic bilinears} $(0<x<y<L)$:
\begin{eqnarray} \label{bosonRL}
\psi_R^{\dag}(x)e^{ie\int_y^x A(x')dx'}\psi_L(y)=
-\frac{1}{L}C_-U_+^{\dag}U_- e^{-2\pi i\frac{x}{L}Q_+
-2\pi i\frac{y}{L}Q_-} e^{-i\pi \frac{x-y}{L}}\\ \nonumber
 \times \exp \sum_{k>0}\left
[\frac{1}{k}+\beta_{x,y}(k)A^{\dag}(k)-\beta^{\star}_{y,x}(k) B^{\dag}(k)
-\beta^{\star}_{x,y}(k)A(k)+\beta_{y,x}(k)B(k)\right]~,
\end{eqnarray}
\begin{eqnarray} \label{bosonRR}
\psi_R^{\dag}(x)e^{ie\int_y^x A(x')dx'}\psi_R(y)=
\frac{1}{L}e^{-2\pi i\frac{x-y}{L}Q_+}e^{i\pi \frac{x-y}{L}}\\ \nonumber
 \times \exp \sum_{k>0}\left[
\frac{1}{k}+\rho_{x,y}(k)A^{\dag}(k)-\sigma^{\star}_{y,x}(k) B^{\dag}(k)
-\rho^{\star}_{x,y}(k)A(k)+\sigma_{y,x}(k)B(k)\right]~,
\end{eqnarray}
\begin{eqnarray} \label{bosonLR}
\psi_L^{\dag}(x)e^{ie\int_y^x A(x')dx'}\psi_R(y)=
-\frac{1}{L}C_-^{\dag}U_-^{\dag}U_+ e^{2\pi i\frac{x}{L}Q_-
+2\pi i\frac{y}{L}Q_+} e^{i\pi \frac{x-y}{L}}\\ \nonumber
 \times \exp \sum_{k>0}\left
[\frac{1}{k}-\beta_{y,x}(k)A^{\dag}(k)+\beta^{\star}_{x,y}(k) B^{\dag}(k)
+\beta^{\star}_{y,x}(k)A(k)-\beta_{x,y}(k)B(k)\right]~,
\end{eqnarray}
and
\begin{eqnarray} \label{bosonLL}
\psi_L^{\dag}(x)e^{ie\int_y^x A(x')dx'}\psi_L(y)=
\frac{1}{L}e^{2\pi i\frac{x-y}{L}Q_-}e^{-i\pi \frac{x-y}{L}}\\ \nonumber
 \times \exp \sum_{k>0}\left
[\frac{1}{k}+\sigma_{x,y}(k)A^{\dag}(k)-\rho^{\star}_{y,x}(k)
B^{\dag}(k)
-\sigma^{\star}_{x,y}(k)A(k)+\rho_{y,x}(k)B(k)\right]~~,
\end{eqnarray}
where
\begin{equation}
\beta_{x,y}(k) \equiv \frac{1}{\sqrt{k}}\left[e^{-2\pi ik \frac{x}{L}}
\cosh\alpha(k) -e^{-2\pi ik \frac{y}{L}} \sinh \alpha(k)
\right]~,
\label{betaxy}
\end{equation}
\begin{equation}
\rho_{x,y}(k) \equiv \frac{1}{\sqrt{k}}\cosh \alpha(k)
\left(e^{-2\pi ik\frac{x}{L}} -e^{-2\pi ik\frac{y}{L}}\right) ~,
\label{rhoxy}
\end{equation}
\begin{equation}
\sigma_{x,y}(k) \equiv \frac{1}{\sqrt{k}}\sinh \alpha(k)
\left(e^{-2\pi ik\frac{x}{L}} -e^{-2\pi ik\frac{y}{L}}\right) ~.
\label{sigmaxy}
\end{equation}

\section{Thermodynamical expectation values. Canonical calculations.}
We want to calculate the thermodynamical expectation value (t.e.v.)
\beeq
\la \ldots \ra_{\beta} = \frac{1}{Z}{\rm Tr}_{\rm phys}\left\{ \ldots
e^{-\beta H}
\right\},
\eneq
where
\beeq
Z={\rm Tr}_{\rm phys} \left(e^{-\beta H}\right)
\eneq
is a partition function. The Trace has to be taken with respect to the
physical, gauge invariant states. Since the total Hamiltonian
(\ref{hamil}) is
a sum of the 'vacuum' Hamiltonian $H_{vac}$ and 'excited' Hamiltonian
$H_{exc}$ and $[H_{vac},H_{exc}] = 0$, the Hilbert space, where the
Hamiltonian $H$ acts, can be expressed as a direct product of the Hilbert
spaces ${\cal H}_{vac}$ (with the states with space momentum $k=0$) and
${\cal H}_{exc}$ (with the states with $k\neq 0$) and we have very
important factorization
\beeq
\la \ldots \ra_{\beta}= \la \ldots \ra_{\beta, vac}\times \la \ldots \ra_
{\beta, exc} ~~~~,
\label{averbeta}
\eneq
where
\beqn
\la \ldots \ra_{\beta, vac} = \frac{1}{Z_{vac}}{\rm Tr}_{{\cal H}_{vac}}
\left \{\ldots e^{-\beta H_{vac}}\right\}, \nonumber \\
Z_{vac}=
{\rm Tr}_{{\cal H}_{vac}}\left( e^{-\beta H_{vac}}\right)
\eeqn
and
\beqn
\la \ldots \ra_{\beta, exc}= \frac{1}{Z_{exc}}{\rm Tr}_{{\cal H}_{exc}}
\left \{\ldots e^{-\beta H_{exc}}\right\}, \nonumber \\
Z_{exc}=
{\rm Tr}_{{\cal H}_{exc}}\left( e^{-\beta H_{exc}} \right).
\eeqn

Let us start with the vacuum sector.\\
Physical, gauge invariant states in ${\cal H}_{vac}$ have a form
(\ref{vgis}). As we know the Hamiltonian $H_{vac}$ has a discrete spectrum
Eq.(\ref{vspec}) and its eigenstates $|E_n \ra $ can be taken as a basis
in
the
space ${\cal H}_{vac}$. So
\beqn
{\rm Tr}_{{\cal H}_{vac}}\left( \ldots e^{-\beta H_{vac}} \right)
=\sum_{n} \la E_n|\ldots|E_n \ra e^{-\beta E_n},
\eeqn
where
\beeq
|E_n \ra = \sum_{N} \int_0^1 d \bar{A} \Psi_{N,n}(\bar{A}) |N,\bar{A} \ra
\label{En}
\eneq
and the wave function $\Psi_{N,n}(\bar{A})$ is a solution Eq.(\ref{Psi})
of
the Schroedinger equation (\ref{SE}).

 For the partition function in the vacuum sector we get
\beqn
Z_{vac}&=& \sum_{n}\la E_n|e^{-\beta H_{vac}}|E_n \ra \nonumber \\
&=& \sum_{N',N,n} \int^1_0 d\bar{A}\int^1_0 d\bar{A}'
\la N',\bar{A}'|\Psi^*_{N',n}(\bar{A'})
\Psi_{N,n}(\bar{A})|N,\bar{A}\ra e^{-\beta E_n} \nonumber \\
&=& e^{-\frac{\beta m}{2}+\frac{\beta\pi}{6L}}
\sum_{N,n}\int^1_0 d\bar{A}\Psi^*_{N,n}(\bar{A})
\Psi_{N,n}(\bar{A})e^{-\beta m n} \nonumber \\
&=& e^{-\frac{\beta m}{2}+\frac{\beta\pi}{6L}}
\left(\frac{\omega}{\pi}\right)^{1/2}
\sum_{N}\int^1_0 d\bar{A}\exp [-\omega(\bar{A} -N+1/2)^2] \nonumber \\
& \times & \sum_n\frac{e^{-\beta  m n}}{2^n n!}
H_n(\sqrt{\omega}(\bar{A} -N+ 1/2))
H_n(\sqrt{\omega}(\bar{A} -N+ 1/2)),
\eeqn
where we have used the orthogonality of the vacuum states
\beqn
\la N',\bar{A'}|N,\bar{A} \ra = \delta_{N',N}\delta(\bar{A'}- \bar{A}).
\eeqn
and the expression Eq.(\ref{vspec}) for the spectrum.
From this with help of the Mehler formula (see e.g. \cite{Bat}):
\begin{eqnarray}
 e^{-\frac{1}{2}x^2} e^{-\frac{1}{2}y^2}\sum_{n=0}^{\infty}2^{-n}H_n(x)H_n(y)
\frac{\xi ^n}{n!}&&\nonumber \\
 =(1-\xi^2)^{-1/2}\exp \left[\frac{4xy\xi - (x^2 +y^2)(1+\xi^2)]}
{2(1-\xi^2)}\right]&& 
\label{Meh}
\end{eqnarray}
and Manton's periodicity conditions (\ref{pc}), which allows the
extension of the integration interval from $[0,1]$ to $( -\infty,
\infty)$ we get
\beqn
Z_{vac}  =
e^{-\frac{\beta m}{2}+\frac{\beta \pi}{6L}}
\left(\frac{\omega}{\pi}\right)^{1/2}
(1- e^{-2\beta m})^{-1/2}
\int_{-\infty}^{\infty}d\bar{A} e^{-\omega \bar{A}^2
\tanh\frac{\beta m}{2}}
\eeqn
and by evaluating the Gaussian integral the final result
\begin{equation}
Z_{vac} = \frac{e^{\frac{\beta \pi}{6L}}}{2 \sinh(\frac{\beta m}{2})}.
\label{vpartf}
\end{equation}
In the same way we can get for the t.e.v. of any gauge invariant
(under all (small and large) gauge transformations)
quantity $F(\bar{A})$
\beqn
\la F(\bar{A})\rab = \la F(\bar{A})\rabv = \sqrt{\frac{\omega
\tanh\frac{\beta m}{2}}{\pi}}\int_{-\infty}^{\infty} d \bar{A} e^{-\omega
\bar{A}^2 \tanh \frac{\beta m}{2}}F(\bar{A})~.
\eeqn
Of course, the result Eq.(\ref{vpartf}) can be obtained just by summation
\beqn
Z_{vac} = \sum_{n=0}^{\infty}e^{-\beta E_n}=
e^{-\frac{\beta m}{2}+\frac{\beta \pi}{6L}}\sum_{n=0}^{\infty}e^{-\beta
mn}
=\frac{e^{\frac{\beta
\pi}{6L}}}{2 \sinh(\frac{\beta m}{2})}.
\eeqn

Now let us consider the excited sector. Using the property of Trace we get
for any operator $O$
\beqn
{\rm Tr}_{{\cal H}_{exc}}\left( Oe^{-\beta H_{exc}} \right)
&=&{\rm Tr}_{{\cal H}_{exc}}\left({\cal U} O{\cal U}^{-1}{\cal U}e^{-\beta
H_{exc}}{\cal U}^{-1} \right) \nonumber \\
&=&{\rm Tr}_{{\cal H}_{exc}}\left( {\tilde O}e^{-\beta {\tilde H}_{exc}}
\right),
\eeqn
where the operators with tilde are those which one obtains after
implementation of the Bogoliubov transformation. Since ${\tilde H}_{exc}$
is just an infinite sum of the Hamiltonians of independent harmonic
oscillators the calculation of t.e.v. becomes a simple task, since
${\tilde O}$ operator will be written in terms of operators $A$ and $B$
and
their Hermitian conjugates. For the calculations one should use
straightforward generalizations of the formulae for one or two harmonic
oscillators given in the Appendix B.

\subsection{Fermionic condensate}
We have
\beqn
\la \bar{\psi}(x)\psi(x)\ra_{\beta} =
\la \psi_R^{\dag}(x)\psi_L(x)\ra_{\beta}+
\la \psi_L^{\dag}(x)\psi_R(x)\ra_{\beta}~.
\eeqn
The vacuum part $\la \psi_R^{\dag}(x)\psi_L(x)\ra_{\beta, vac}$
is essentially calculated like $Z_{vac}$. Using the bosonization formula
(\ref{bosonf}) we get

\begin{eqnarray}
&& {\rm Tr}_{{\cal H}_{vac}} \{\psi_R^{\dag}(x)\psi_L(x)e^{-\beta H_{vac}}
\}= -\frac{1}{L}
 \sum_{N,N',n}\int^1_0 d\bar{A} \int ^1_0 d\bar{A}' \\
& \times &
\Psi^*_{N',n}(\bar{A}')\la N',\bar{A}'|U_+^{\dagger}U_-|N,\bar{A} \ra
e^{-\beta E_n }\Psi_{N,n}(\bar{A})  \nonumber \\
&= &-\frac{1}{L}\left(\frac{\omega}{\pi}\right)^{1/2}
e^{-\frac{\beta m}{2}+\frac{\beta \pi}{6L}}
\sum_{N',N,n} \int^1_0
d\bar{A}\delta_{N',N+1} \nonumber \\
&\times&\exp [-(\omega/2)(\bar{A} -N'+1/2)^2]
\exp [-(\omega/2)(\bar{A} -N+1/2)^2] \nonumber \\
&\times&\sum_n\frac{e^{-\beta m n}}{2^n n!}
H_n(\sqrt{\omega}(\bar{A} -N'+ 1/2))
H_n(\sqrt{\omega}(\bar{A} -N+ 1/2)) \nonumber \\
& = &-\frac{1}{L}\left(\frac{\omega}{\pi}\right)^{1/2}
e^{-\frac{\beta m}{2}+\frac{\beta \pi}{6L}}
(1- e^{-2\beta m})^{-1/2}
\int_{-\infty}^{\infty}d\bar{A} e^{-\omega \bar{A}^2
\tanh(\frac{\beta m}{2})}e^{-\frac{\omega}{4}
\coth(\frac{\beta m}{2})}, \nonumber
\end{eqnarray}
where we have used the equation
\beqn
\la N',\bar{A}'|U_+^{\dag}U_-|N,\bar{A} \ra = \delta_{N',N+1}
\delta(\bar{A}' -\bar{A}) 
\label{matU},
\eeqn
which follows from the properties of the $U_+$ and $U_-$ operators (see
Appendix A).\\
 Calculating the Gaussian integral and using the result Eq.(\ref{vpartf}),
we
get finally
the vacuum part
\beqn
\la \psi_R^{\dag}(x)\psi_L(x)\ra _{\beta, vac} =
- \frac{1}{L}e^{-\frac{\pi}{m L}\coth (\frac{\beta m}{2})}.
\label{chcvac}
\eeqn
In order to calculate the t.e.v. over the
states with space momentum
$k\neq 0$, we have to use the bosonization formula (\ref{bosonf})
and the formula (\ref{tho2}) from Appendix B.
\begin{eqnarray}
&&\la \psi_R^{\dag}(x)\psi_L(x)\ra_{\beta, exc} \nonumber \\
&&=\la \exp \sum_{k>0}\left[ \frac{1}{k}
+\beta_x(k)
A^{\dagger}(k)-\beta_x^{\ast}(k)B^{\dagger}(k)
-\beta_x^{\ast}(k)A(k)+\beta_x(k)B(k)\right] \ra_{\beta, exc}
\nonumber
\\
&&= \exp\sum_{k>0}\left (
\frac{1}{k}
- \beta_x^{\ast}(k)\beta_x(k) \coth \frac{\beta E(k)}{2} \right
).
\end{eqnarray}
According to Eqs.(\ref{betax}) and (\ref{alpha}):
\begin{equation}
\beta_x^{\ast}(k)\beta_x(k) = \frac{1}{k}(\cosh \alpha(k) - \sinh
\alpha(k) )^2 = \frac{2\pi}{E(k)L}.
\end{equation}
So we get our final result
\begin{equation}
\la \psi_R^{\dag}(x)\psi_L(x)\ra_{\beta} =
- \frac{1}{L}e^{-\frac{\pi}{mL}
\coth (\frac{\beta m}{2})}
e^{\sum_{k >0}\{ \frac{1}{k} - \frac{2\pi}{LE(k)}\coth (\frac{\beta
E(k)}{2})\}}~.
\label{chcond}
\end{equation}
The same result we will get for $\la \psi_L^{\dag}(x) \psi_R(x)
\ra_{\beta}$ if we use the fact that
\begin{displaymath}
\la N',\bar{A}'|U_-^{\dag}U_+|N,\bar{A} \ra = \delta_{N',N-1}
\delta(\bar{A}'-\bar{A}).
\end{displaymath}
So
\beqn
\la \bar{\psi}(x)\psi(x)\ra_{\beta} =
- \frac{2}{L}e^{-\frac{\pi}{mL}
\coth (\frac{\beta m}{2})}
e^{\sum_{k >0}\{ \frac{1}{k} - \frac{2\pi}{LE(k)}\coth (\frac{\beta
E(k)}{2})\}}~.
\label{fermcond2}
\eeqn
In path integral formulation  of the SM on a torus we have  the
following results for the chiral fermionic condensate \cite {JA}, \cite{SW}, \cite{Azakov}, \cite{Azakov2}:
\begin{eqnarray}
&&\langle \bar{\psi}(x)P_{\pm} \psi (x)\rangle =
-\frac{\eta^2(\tau)}{L_1}e^{2e^2 G(0) -\frac{2\pi ^2}{e^2L_1L_2}}
\label{cfc}
\end{eqnarray}
where $P_{\pm}=\frac{1}{2}(1\pm \gamma_{5})$, $L_1$ and $L_2$ are lengths of 
two circumferences of a torus, $\eta(\tau)$ is Dedekind's function \cite{theta}, \cite{Bat} and $\tau = i \frac{L_2}{L_1}$. \\
The propagator $G(x)$ satisfies the following equation
\begin{equation}
\Box(\Box -m^2)G(x-y)
 = \delta^{(2)}(x-y) -\frac{1}{L_1L_2}~, \label{prop11}
\end{equation}
where $\delta^{(2)}(x-y)$ is Dirac's $\delta$-function on the torus.\\
It can be written as the difference of a massless and massive propagator
on the torus orthogonal to the constant functions:
$   G(x)=  1/m^2\{ G_0(x) - G_m(x)\}$. There is a closed expression in
the massless case written through Jacobi's $\theta$ funcions \cite{theta}, \cite{Bat}:
\begin{equation} \label{zprop 1}
 G_0(x)= -\frac{1}{2\pi} \log \left( 2 \pi \eta^2(\tau)
   e^{-\pi\frac{x_2^2}{L_1 L_2}}
 \frac{|\theta_1(z|\tau)|}{|\theta_1'(0|\tau)|}\right)~,
\end{equation}
where $z=\frac{x_1+ix_2}{L_1}$.
It can also be written as the infinite sum
\begin{equation}
G_0(x)=\frac{1}{4\pi}\sum_{n\neq 0}\frac{1}{n}\frac{\cosh \big[\frac{2\pi
n}
{L_1}\big(\frac{L_2}{2}-|x_2| \big)\big]}{\sinh (\pi n|\tau|)}e^{2\pi
in\frac{x_1}{L_1}} -\frac{|x_2|}{2L_1}+\frac{x_2^2}{2L_1L_2}+\frac{|\tau|}
{12}~. \label{zprop2}
\end{equation}
In the massive case we use the infinite sum for
$\overline{G}_m(x)= G_m(x) + 1/m^2 L_1 L_2 $:
 \begin{eqnarray}
&&\overline{G}_m(x)=\frac{1}{2 L_1 }\sum_n
\frac{\cosh[ E(n)(L_2/2 - | x_2 |)]
e^{2\pi in \frac{x_1}{L_1}}} {E(n)\sinh [L_2 E (n)/2]},\nonumber \\
&&=\frac{1}{2L_1m}\Big (\coth \frac{mL_2}{2} \cosh m|x_2| -\sinh m|x_2|
\Big) \nonumber \\
&&+\sum_{n>0} \frac{\cos \left (2\pi n \frac{x_1}{L_1} \right)}{L_1E(n)}
\left [\coth \left(\frac{E(n)L_2}{2} \right) \cosh (E(n)|x_2|)
-\sinh(E(n)|x_2|) \right]~,
\label{mprop}\\
&&E(n) = \left[ \frac{4 \pi^2n^2}{L_1^2} + m^2\right]^{1/2}~.
\nonumber
\end{eqnarray}
From Eqs (\ref{mprop}) and (\ref{zprop2}) we see that $G_0(x)$ is a
limiting case of $G_m(x)$ when $m \rightarrow 0$.
With the help of the equations (\ref{zprop2}) and (\ref{mprop}) the expression for chiral condnsate Eq.(\ref{cfc})
can be rewritten
\begin{eqnarray}
&&\langle \bar{\psi}(x)P_{\pm} \psi 
(x)\rangle=-\frac{1}{L_1}e^{-\frac{\pi}{L_1m}\coth \frac{mL_2}{2}}
e^{\sum_{n>
0}\left\{\frac{1}{n} -\frac{2\pi}{L_1E(n)}
\coth \left(\frac{E(n)L_2}{2} \right) \right \}}.
 \label{cfc1}
\end{eqnarray}
which is exactly our equation (\ref{chcond}) if we put in it $\beta = L_2 
,~~L=L_1$.
\subsection{Currents correlation function}
Using Eq.(\ref{currx}) we get
\beqn
\la j_+(x,t)j_+(x',t') \rab =\frac{1}{L^2} \la Q_+(t)Q_+(t') \rabv \cr
+\frac{1}{L^2} \sum_{k\not=0,k'\not=0}
e^{2\pi i k \frac{x}{L} +2\pi i k' \frac{x'}{L}}
\la j_+(k,t) j_+(k',t') \rabe .
\label{tevcur}
\eeqn

Let us first calculate $\la Q_+(t)Q_+(t') \rabv$ . On the physical space
$Q_+= -Q_- = Q_5/2$ and the vacuum Hamiltonian has a form (we omit the
constant term $-\frac{\pi}{6L}$ which is inessential for the calculations
of the expectation values)
\beeq
H_{vac} = \frac{L}{2} F^2 + \frac{\pi}{2L} Q_5^2. \label{hvac1}
\eneq
Using a commutation relation:
\beeq
[F, Q_5] = i \frac{e}{\pi} ,
\eneq
we can introduce creation $(a^{\dag})$ and annihilation $(a)$ operators:
\beqn
a^{\dag} \equiv  \sqrt{\frac{\omega}{2}} \left( \frac{Q_5}{2}
+i\frac{L}{2\sqrt{\pi}} F \right), \nonumber\\
a \equiv \sqrt{\frac{\omega}{2}} \left( \frac{Q_5}{2}
-i\frac{L}{2\sqrt{\pi}} F \right), \label{caoper}
\eeqn
which obey the canonical commutation relation
\[ [a,a^{\dag}] = 1 .  \]
Then the Hamiltonian Eq.(\ref{hvac1}) takes a form
\beeq
H_{vac}= m(a^{\dag}a +\frac{1}{2} ), \label{hvac2}
\eneq
and with the help of Eq.(\ref{caoper}) and Eq.(\ref{tho3}) we
obtain the time dependence of the axial charge:
\beeq
Q_5(t) =e^{itH_{vac}}Q_5 e^{-itH_{vac}} = \sqrt{\frac{2}{\omega}}
\left( e^{imt}a^{\dag} +e^{-imt}a \right).
\eneq
Now using the formulae given in the Appendix B
we can easily calculate t.e.v.
\beqn
\la Q_5(t)Q_5(t') \rabv & = & \frac{2}{\omega} \left [ e^{im(t-t')}
\la a^{\dag}a \rabv + e^{-im(t-t')} \la a a^{\dag} \rabv \right] \cr
& = & \frac{2}{\omega} \frac {\cosh m(\frac{\beta}{2}-i(t-t'))}
{\sinh \frac{m\beta}{2} }.
\eeqn
So
\beqn
&&\la Q_+(t)Q_+(t') \rabv = \la Q_-(t)Q_-(t') \rabv= - \la Q_+(t)Q_-(t')
\rabv \cr
&&= -\la Q_-(t)Q_+(t') \rabv = \frac{mL}{8\pi} \frac{ \cosh
m(\frac{\beta}{2}
-i(t-t'))}{\sinh \frac{m\beta }{2}}. 
\label{tevchar}
\eeqn

Now let us calculate $\la j_+(k,t) j_+(k',t') \rabe $.~
To this aim  we will use the expression of currents in terms of
operators $A, A^{\dag},B $ and $B^{\dag}$ given in Eqs.(\ref{currk}).
The nonzero contribution comes from the terms, where $k$ and $k'$ have
opposite signs.
E.g. in the case, where $ k>0, k'<0$ :
\beqn
\la j_+(k,t) j_+(k',t') \rabe  = k\delta _{k,-k'}
\left [ \la A(k,t)A^{\dag}(k,t') \rabe \cosh^2\alpha(k)\right. \nonumber\\
+ \left.\la B^{\dag}(k,t) B(k,t') \rabe \sinh^2\alpha(k) \right] ~.
\label{tevcurk}
\eeqn
Then we have
\[ A(k,t) = e^{itH_{exc}}A(k)e^{-itH_{exc}}= e^{-iE(k)t}A(k)~, \]
\[ A^{\dag}(k,t) = e^{itH_{exc}}A^{\dag}(k)e^{-itH_{exc}}= e^{iE(k)t}
A^{\dag}(k)~,
\]
\[ \la A(k)A^{\dag}(k) \rabe = \frac{1}{1 - e^{-\beta E(k)}}~, \]

\[ \la A(k)^{\dag} A(k) \rabe = \frac{-1}{1 - e^{\beta E(k)}}~, \]
and the same for $B$ operators. Following this way and using
Eq.(\ref{alpha}) we finally get
\beqn
&&\la j_+(x,t) j_+(x',t') \rabe \cr
&&=\frac{1}{i}(\pardx -\pardt )\frac{1}{4\pi L}\sum_{k\neq 0}
\frac{e^{2\pi
ik\frac{\bar{x}}{L}}}{\sinh \frac{\beta E(k)}{2}} \sinh \left[
E(k)(\frac{\beta}{2} - i\bar{t})  \right ] \cr
&& -\frac{e^2}{8\pi^2L} \sum_{k\neq 0}  \frac{e^{2\pi
ik\frac{\bar{x}}{L}}}{2E(k)\sinh \frac{\beta E(k)}{2}} \cosh \left[
E(k)(\frac{\beta}{2} - i\bar{t})  \right ] , 
\label{tevcur1}
\eeqn
where $ \bar{x} \equiv x-x', \bar{t}\equiv t-t' $.
\par
We have the following t.e.v. in the {\bf theory of the free quantum
neutral massive ($m=e/\sqrt{\pi}$) scalar field $A(x,t)$ on the circle}
($x \neq x', t \neq t'$):
\beqn
\la A(x,t)A(x',t') \rab &=&\frac{1}{2L}\sum_k  \frac{e^{2\pi
ik\frac{\bar{x}}{L}}}{E(k)\sinh \frac{\beta E(k)}{2}} \cosh \left[
E(k)(\frac{\beta}{2} - i\bar{t})  \right ] ,\cr
&\equiv& \gbm = \ngbm + \tgbm, \label{tevmsf}
\eeqn
where $\ngbm $ is a term with $k=0$ :
\beqn
\ngbm \equiv \frac{1}{2L}\frac{\cosh \left[m(\frac{\beta}{2}-i\bar{t})\right]}
{m \sinh \frac{\beta m}{2}}\label{prop0}
\eeqn
and $\tgbm $ is the rest:
\beqn
\tgbm \equiv \frac{1}{2L}\sum_{k\neq 0}  \frac{e^{2\pi
ik\frac{\bar{x}}{L}}}{E(k)\sinh \frac{\beta E(k)}{2}} \cosh \left[
E(k)(\frac{\beta}{2} - i\bar{t})  \right ] \label{proptil}~.
\eeqn
Note that $\gbm$ obeys the equation
\beeq
m^2 \gbm = \left( -\pardt^2 + \pardx^2 \right) \gbm~. \label{eqGbeta}
\eneq
Then from Eqs.(\ref{tevcur1}) and (\ref{tevchar}) we get
\beeq
\la j_+(x,t)j_+(x',t') \rabe = \frac{1}{2\pi}\left( \pardx \pardt -
\pardt^2 - \frac{m^2}{2} \right) \tgbm
\eneq
and
\beeq
\frac{1}{L^2} \la Q_+(t)Q_+(t') \rabv = \frac{1}{2\pi}\frac{m^2}{2}
\ngbm~.
\eneq
Finally
\beqn
\la j_+(x,t)j_+(x',t') \rab & = & \frac{1}{2\pi}\left( \pardx \pardt
-\pardt^2 - \frac{m^2}{2} \right) \gbm \cr
&=&\frac{1}{4\pi}\left ( -\pardx^2 -\pardt^2 +2\pardx \pardt \right) \gbm
~,
\eeqn
where in order to get the second line we used Eq.(\ref{eqGbeta}).\\
Similarly it can be shown that
\beeq
\la j_-(x,t)j_-(x',t') \rab  = \frac{1}{4\pi}\left ( -\pardx^2 -\pardt^2
-2\pardx \pardt \right) \gbm
\eneq
and
\beeq
\la j_+(x,t)j_-(x',t') \rab = \la j_-(x,t)j_+(x',t') \rab =
-\frac{1}{4\pi} m^2 \gbm
\eneq
Now since $j_{\pm}(x,t) = \frac{1}{2}(j_0(x,t)\pm j_1(x,t))$ we will get
\beeq
\la j_0(x,t)j_0(x',t') \rab = -\frac{1}{\pi}\pardx^2 \gbm~,
\label{tevcurx0}
\eneq

\beeq
\la j_1(x,t)j_1(x',t') \rab = - \frac{1}{\pi}\pardt^2 \gbm~,
\label{tevcurx1}
\eneq

\beeq
\la j_0(x,t)j_1(x',t') \rab = \frac{1}{\pi} \pardx \pardt \gbm~.
\label{tevcurx01}
\eneq
Thus
\beeq
\la j_{\mu}(x,t)j_{\nu}(x',t') \rab =  -\frac{1}{\pi} \varepsilon_{\mu
\rho } \varepsilon_{\nu \sigma} \partial_{\rho} \partial_{\sigma} \gbm ~.
\label{tevcurx}
\eneq

%%%%%%%%%%%%%%%%%%%%%%%%%%%%%%%%%%%%%%%%%%%%%%%%%%%%%%
\subsection{Correlation function of the electric fields}
Now let us calculate the t.e.v.:$\la F(x,t)F(x',t')\rab$, where
$F(x,t) = e^{iHt}F(x)e^{-iHt}$ and $F(x)$ is the electric field.
From Eq.(\ref{elecf}) we get (for $x\neq x', t\neq t'$)
\beqn
\la F(x,t)F(x,t) \rab &=&\la F(t)F(t')\rabv \nonumber\\
&+& \frac{1}{L^2}\sum_{k\neq 0}\sum_{k'\neq 0}
\la f(k,t)f(k',t')\rabe e^{2\pi i(k\frac{x}{L} +k'\frac{x'}{L})}
\eeqn
With the help of Eq.(\ref{GL}) the calculation of
$\la f(k,t)f(k',t') \rabe $
is reduced to the calculation of t.e.v. of currents which was done before
(see e.g. Eq.(\ref{tevcurk})). So we obtain
\beqn
\la F(x,t)F(x',t') \rab = \la F(t)F(t') \rabv + m^2\tgbm~~,
\eeqn
where  $\tgbm$ is given in Eq.(\ref{proptil}). From Eqs.(\ref{caoper}) and
(\ref{hvac2}) it follows that
\beqn
\la F(t)F(t') \rabv = m^2 \ngbm~~,
\eeqn
where $\ngbm$ is defined in  Eq.(\ref{prop0}). Finally we get using
Eqs.(\ref{tevmsf}) and (\ref{eqGbeta}):
\beqn
\la F(x,t)F(x',t') \rab = (-\pardt^2+\pardx^2) \gbm
\label{corelecf}
\eeqn
%%%%%%%%%%%%%%%%%%%%%%%%%%%%%%%%%%%%%%%%%%%%%%%%%%%%%%%%%%%%%%%%%%%%%%%%%%%%

\subsection{Thermodynamical expectation values of the gauge invariant
fermion bilinears}

From Eq.(\ref{bosonRL}) using Eq.(\ref{tho2}) we obtain in the excited
sector
\beqn
\la \psid_R(x)e^{ie\int_y^x A(x')dx'}\psi_L(y) \rabe
=e^{\sum_{k>0}\left [\frac{1}{k}-\frac{1}{2}\left(|\beta_{x,y}(k)|^2
+|\beta_{y,x}(k)|^2\right)\coth\frac{\beta E(k)}{2} \right ]}.
\eeqn
Using the explicit form of $\beta_{x,y}$ given in Eq.(\ref{betaxy}) and
Eq.(\ref{alpha}) we get
\beqn
|\beta_{x,y}(k)|^2 +|\beta_{y,x}(k)|^2 =
\frac{2}{k}\left[\cosh2\alpha(k) -\sinh2\alpha(k)
\cos2\pi k \frac{x-y}{L} \right] \nonumber\\
=2\left[\frac{1}{E(k)}\left(\frac{2\pi}{L}+\frac{e^2L}{4\pi^2k^2}\right)
-\frac{e^2L}{4\pi^2E(k)k^2}\cos2\pi k\frac{x-y}{L} \right]~.
\eeqn
So
\beqn
\la \psid_R(x)e^{ie\int_y^x A(x')dx'}\psi_L(y) \rabe
=\exp\left\{\sum_{k>0}\left
[\frac{1}{k}-\frac{2\pi}{LE(k)}\coth\frac{\beta
E(k)}{2}  \right ]+I(x-y)\right\}.
\eeqn
where
\beqn
I(x)\equiv
-\frac{e^2L}{4\pi^2}\sum_{k>0}\frac{\coth\frac{\beta E(k)}{2}}{E(k)k^2}
\left(1-\cos2\pi k\frac{x}{L}\right).
\label{funcI}
\eeqn
The calculations of the expectation value in the vacuum sector are similar
to those of chiral condensate. From Eq.(\ref{bosonRL}) we have
\beqn
\la \psid_R(x)e^{ie\int_y^x A(x')dx'}\psi_L(y) \rabv
=-\frac{1}{L}\la U_+^{\dag}U_-e^{-2\pi i \frac{(x-y)}{L}Q_+}\rabv e^{-i\pi
\frac{x-y}{L}}~~.
\eeqn
From Eq.(\ref{En}) it follows that
\beqn
e^{-2\pi i \frac{(x-y)}{L}Q_+}|\kE = \sum_N \int_0^1 d \bar{A}
\Psi_{N,n}(\bar{A})e^{-2\pi i \frac{(x-y)}{L}(N-\bar{A}- 1/2)}|N,\bar{A}
\ra~~.
\eeqn
Then from Eq.(\ref{matU}) we get
\beqn
\bE|U_+^{\dag}U_-e^{-2\pi i \frac{(x-y)}{L}Q_+}|\kE =\sum_N \int_0^1 d
\bar{A} \Psi_{N+1,n}^*(\bar{A})
\Psi_{N,n}(\bar{A})e^{-2\pi i \frac{(x-y)}{L}(N-\bar{A}- 1/2)}
\eeqn
and from Eq.(\ref{Psi}) and the Mehler formula (\ref{Meh}) it follows
\beqn
&&\sum_n \bE|U_+^{\dag}U_-e^{-2\pi i \frac{(x-y)}{L}Q_+}|\kE e^{-\beta
E_n}
=\left(\frac{\omega}{\pi}\right)^{1/2}
e^{-\frac{\beta m}{2}+\frac{\beta \pi}{6L}}
(1- e^{-2\beta m})^{-1/2}e^{-\frac{\omega}{4}
\coth \frac{\beta m}{2}} \nonumber\\
&&\times \sum_N \int_0^1 d
\bar{A} e^{-2\pi i \frac{(x-y)}{L}(N-\bar{A}- 1/2)}
e^{-\omega(N-\bar{A})^2 \tanh\frac{\beta m}{2}}~~.
\eeqn
Now again extending the integration from [0,1] to $(-\infty, \infty)$  we
obtain that
\beqn
\la U_+^{\dag}U_-e^{-2\pi i \frac{(x-y)}{L}Q_+}\rabv =
e^{-\frac{\omega}{4}\coth \frac{\beta m}{2}}e^{\pi i\frac{x-y}{L}}
e^{-\frac{\pi^2(x-y)^2}{\omega L^2}\coth \frac{\beta m}{2}}
\eeqn
and finally taking into account Eq.(\ref{chcond})
\beqn
\la \psid_R(x)e^{ie\int_y^x A(x')dx'}\psi_L(y) \rab =
\la \psid_R \psi_L \rab e^{-\frac{\pi m}{4L}(x-y)^2\coth \frac{\beta m}{2}
+I(x-y)}~~.
\eeqn
The same result we will obtain for $\la \psid_L(x)e^{ie\int_y^x
A(x')dx'}\psi_R(y) \rab$.
\par
Now let us calculate $ \la \psid_R(x)e^{ie\int_y^x A(x')dx'}\psi_R(y)
\rab$. Using Eq.(\ref{tho2}) once more we obtain from Eq.(\ref{bosonRR}):
\beqn
\la \psid_R(x)e^{ie\int_y^x A(x')dx'}\psi_R(y)\rabe  \nonumber
=e^{\sum_{k>0}\left [\frac{1}{k}-\frac{1}{2}\left(|\rho_{x,y}(k)|^2
+|\sigma_{y,x}(k)|^2\right)\coth\frac{\beta E(k)}{2} \right ]}.
\eeqn
From Eqs.(\ref{rhoxy}), (\ref{sigmaxy}) and (\ref{alpha}) it follows that
\beqn
|\rho_{x,y}(k)|^2 +|\sigma_{y,x}(k)|^2 = \frac{2}{E(k)}
\left( \frac{2\pi}{L}+\frac{e^2L}{4\pi^2 k^2} \right)
\left[1-\cos 2\pi k\frac{(x-y)}{L}\right]
\eeqn
and
\beqn
\la \psid_R(x)e^{ie\int_y^x A(x')dx'}\psi_R(y) \rabe =
\la \psid_R \psi_L \rab e^{2\pi
\bar{G}_{\beta,m}(x-y,0)+I(x-y)}~~,
\eeqn
where in the last step we used definitions of
$\bar{G}_{\beta,m}(x,0)$ and $I(x)$ given in Eqs.(\ref{tevmsf})-
(\ref{proptil}) and
Eq.(\ref{funcI}), respectively and Eq.(\ref{chcond}).

In the vacuum sector from Eq.(\ref{bosonRR}) we obtain
\beqn
\la \psid_R(x)e^{ie\int_y^x A(x')dx'}\psi_R(y) \rabv
=\frac{1}{L}\la e^{-2\pi i \frac{(x-y)}{L}Q_+}\rabv e^{i\pi
\frac{x-y}{L}}~~,
\eeqn
where
\beqn
\la e^{-2\pi i \frac{(x-y)}{L}Q_+}\rabv =\frac{1}{Z_{vac}}
\sum_n \bE|e^{-2\pi i \frac{(x-y)}{L}Q_+}|\kE e^{-\beta E_n} \nonumber\\
 =\frac{1}{Z_{vac}}\sum_N
\int_0^1 d\bar{A} e^{-2\pi i \frac{(x-y)}{L}(N-\bar{A}- 1/2)}
\sum_n e^{-\beta E_n}\Psi_{N,n}^*(\bar{A})\Psi_{N,n}(\bar{A})~~.
\eeqn
Again using the Mehler formula (\ref{Meh}) and extending the integration
interval to $(-\infty, \infty)$ we get
\beqn
\la e^{-2\pi i \frac{(x-y)}{L}Q_+}\rabv = e^{-\frac{\pi m}{4L}(x-y)^2
\coth\frac{\beta m}{2}}~~~.
\eeqn
Finally
\beqn
&&\la \psid_R(x)e^{ie\int_y^x A(x')dx'}\psi_R(y) \rab   \nonumber \\
&&=\la \psid_R \psi_L \rab e^{ \pi i \frac{x-y}{L}+2\pi
\bar{G}_{\beta,m}(x-y,0)
-\frac{\pi m}{4L}(x-y)^2
\coth\frac{\beta m}{2}+I(x-y)}
\eeqn
Similarly we will get from Eq.(\ref{bosonLL}):
\beqn
&&\la \psid_L(x)e^{ie\int_y^x A(x')dx'}\psi_L(y) \rab   \nonumber \\
&&=\la \psid_R \psi_L \rab e^{- \pi i \frac{x-y}{L}+2\pi
\bar{G}_{\beta,m}(x-y,0)
-\frac{\pi m}{4L}(x-y)^2
\coth\frac{\beta m}{2}+I(x-y)}
\eeqn

%%%%%%%%%%%%%%%%%%%%%%%%%%%%%%%%%%%%%%%%%%%%%%%%%%%%%%%%%%%%%%%%%%%%%%%%%%%5

\subsection{Densities correlation function  
$\la\bar{\psi}(x,t)\psi(x,t)\bar{\psi}(0,0)\psi(0,0) \ra_{\beta} $}
The t.e.v.
$\la \psibar(x,t)\psi(x,t)\psibar(0,0)\psi(0,0)\ra_{\beta}$ is a sum of
four t.e.v.
$\la(1)\ra_{\beta}, \la(2)\ra_{\beta}, \la(3)\ra_{\beta}$ and
$\la(4)\ra_{\beta}$, where
\beqn
(1) &\equiv& \psid_R(x,t)\psi_L(x,t)\psid_R(0,0)\psi_L(0,0)~,\cr
(2) &\equiv& \psid_L(x,t)\psi_R(x,t)\psid_R(0,0)\psi_L(0,0)~,\cr
(3) &\equiv& \psid_R(x,t)\psi_L(x,t)\psid_L(0,0)\psi_R(0,0)~,\cr
(4) &\equiv& \psid_L(x,t)\psi_R(x,t)\psid_L(0,0)\psi_R(0,0).
\eeqn
The dependence of $t$ comes from:
\beqn
\psid _R(x,t)\psi _L(x,t) = e^{itH}\psid _R(x)\psi_L(x)e^{-itH}~,
\eeqn
where $H$ is the Hamiltonian
\beqn
H=H_{vac}+H_{exc}~.
\eeqn

E.g. for $\la (1)\ra_{\beta}$ we have
\beqn
&&\la\psid_R(x,t)\psi_L(x,t)\psid_R(0,0)\psi_L(0,0)\ra_{\beta, vac}\cr
&&=Z_{vac}^{-1}{\rm Tr}_{{\cal H}_{vac}}\left
\{e^{itH_{vac}}\psid_R(x)\psi_L(x)e^{-itH_{vac}}
\psid_R(0)\psi_L(0)e^{-\beta H_{vac}}\right \}
\eeqn
and
\beqn
&&\la\psid_R(x,t)\psi_L(x,t)\psid_R(0,0)\psi_L(0,0)\ra_{\beta, exc}\cr
&&=Z_{exc}^{-1}{\rm Tr}\left
\{e^{itH_{exc}}\psid_R(x)\psi_L(x)e^{-itH_{exc}}
\psid_R(0)\psi_L(0)e^{-\beta H_{exc}} \right \}
\eeqn

For the calculation of these averages we will use that part of
$\psid_R(x)\psi_L(x)$ which contributes.

From bosonization formula (\ref{bosonf}) it follows that for the
calculation of
$\la\cdots\ra_{\beta, exc}$ we may use instead of $\psid_R(x)\psi_L(x)$
the
following expression
\beqn
e^{\sum_{k>0}\left
\{\frac{1}{k}+\beta_x(k)A^{\dag}(k)-\beta^{\star}_x(k)B^{\dag}(k)
-\beta^{\star}_x(k)A(k)+\beta_x(k)B(k)\right\}}
\eeqn
Now we will use the formula (\ref{tho4}) from Appendix B with the result:
\beqn
&&\langle (n)\rangle_{\beta, exc} = \exp\left \{2\sum_{k>0}\left
(\frac{1}{k}-\frac{2\pi}{LE(k)}
\coth{\frac{\beta E(k)}{2}} \right )\right \} \cr
&&\times\exp\left \{\zeta 4\pi\sum_{k>0}\frac{\cos\left(2\pi
k\frac{x}{L}\right)}{LE(k)}\left [\coth
{\frac{\beta E(k)}{2}}\cosh (it E(k))-\sinh (it E(k)) \right] \right
\},
\label{nexc}
\eeqn
where $\zeta = -1$ for $n=1,4$ and $\zeta =1$ for $n=2,3$.

Now we will prove that
\beqn
\langle(n)\rangle_{\beta, vac}=\frac{1}{L^2}e^{-\frac{2\pi}{Lm}\coth
\frac{m\beta}{2}}
e^{\zeta \frac{2\pi}{Lm}
\left(\coth \frac{m\beta}{2}\cosh(it m))-
\sinh(it m)\right)}~~, 
\label{nvac}
\eeqn
Let us calculate e.g. $\langle(1)\rangle_{\beta, vac}$ (for other $(n)$'s
the
calculations are the same)
\beqn
&&\la\psid_R(x,t)\psi_L(x,t)\psid_R(0,0)\psi_L(0,0)\ra_{\beta, vac}\cr
&&=Z_{vac}^{-1}{\rm Tr}_{{\cal H}_{vac}}\left
\{e^{itH_{vac}}\psid_R(x)\psi_L(x)e^{-itH_{vac}}
\psid_R(0,0)\psi_L(0,0)e^{-\beta H_{vac}}\right \}\cr
&&=Z^{-1}_{vac}\sum_{n,n'}e^{(it -\beta)n\mu}e^{-it n'\mu}
\bE| \psid_R(x)\psi_L(x)|\kEp \bEp| \psid_R(0)\psi_L(0)|\kE~~,
\eeqn
where we used the completeness of the states $\{ |\kE \}$ in the space
${\cal H}_{vac}$
\beqn
\sum_n|\kE \bE| =1.
\eeqn

As a spectrum of the Hamiltonian $H_{vac}$ we take $\{ nm, n=0,1,\ldots
\}$, because the constant terms in the spectrum (\ref{vspec}) are not
important for the t.e.v..\\
Again using (\ref{vspec}) and the bosonization formula (\ref{bosonf}) we
shall get
\beqn
&&\la (1) \ra_{\beta,vac} = Z_{vac}^{-1}\sum_{n,n'}
e^{(it -\beta)nm}e^{-it n'm}\frac{1}{L^2} \cr
&&\times
\sum_{N,N'}
\int_0^1d\bar{A}\int_0^1d\bar{A}'
\Psi^*_{N',n}(\bar{A}')\Psi_{N,n'}(\bar{A})
 \langle N',\bar{A}'|U_+^{\dag}U_-|N,\bar{A}\rangle  \cr
&&\times
\sum_{N_1,N'_1}
\int_0^1d\bar{A_1}\int_0^1d\bar{A_1}'
\Psi^*_{N'_1,n'}(\bar{A}'_1)\Psi_{N_1,n}(\bar{A}_1)
\langle N'_1,\bar{A}'_1|U_+^{\dag}U_-|N_1,\bar{A}_1\rangle
\eeqn
and with the help of (\ref{matU})
\beqn
&&\langle (1)\rangle_{vac}=Z^{-1}_{vac}\sum_{n,n'}e^{(it
-\beta)nm}e^{-it n'm} \cr
&&\times \frac{1}{L^2}\int_0^1d\bar{A}\int_0^1d\bar{A}_1
\sum_{N,N_1}
\Psi^*_{N+1,n}(\bar{A})\Psi_{N,n'}(\bar{A})
\Psi^*_{N_1+1,n'}(\bar{A}_1)\Psi_{N_1,n}(\bar{A}_1). 
\label{(1)}
\eeqn
Using the Mehler's formula (\ref{Meh}) twice in order to do the summations
with respect to $n$ and $n'$ and using again Manton's periodicity
condition
\[ \sum_N \int_0^1d\bar{A} \rightarrow \int_{-\infty}^{\infty}d \bar{A}\]
we obtain (\ref{nvac}). The details of the calculations  are given in the
Appendix C.

From Eqs (\ref{nexc}),(\ref{nvac}),(\ref{fermcond2}) and definitions
(\ref{tevmsf}),(\ref{prop0}),(\ref{proptil}) we obtain
\beqn
\la(n) \ra =\la(n)\ra_{\beta,vac}\la(n)\ra_{\beta,exc}
=\frac{1}{4}(\la \bar{\psi}(x)
\psi(x)\ra_{\beta})^2e^{\zeta 4\pi\bar{G}_{\beta,m}(x,t)}
\eeqn
and
\beqn
\la \bar{\psi}(x,t)\psi(x,t)\bar{\psi}(0)\psi(0) \ra_{\beta} =
(\la \bar{\psi}(x)
\psi(x)\ra_{\beta})^2 \cosh { 4\pi\bar{G}_{\beta,m}(x,t)}
\eeqn

%%%%%%%%%%%%%%%%%%%%%%%%%%%%%%%%%%%%%%%%%%%%%%%%%%%%%%%%%%%%%%%%%%%%%%%%%
\subsection{Correlation function for the product of $n$-chiral scalars}
We introduce chiral scalars
\beqn
S_{\zeta}(x)\equiv \psi^{\dag}_{\zeta}(x) \psi_{-\zeta}(x)~~,
\label{chsca}
\eeqn
where $\zeta =\pm 1, \psi_{+1}(x) \equiv \psi_{+}(x) \equiv \psi_L(x)$
and our aim is to calculate the t.e.v. of the product of $n$ such operators 
at arbitrary times. We will prove the following formula:
\beqn
\la \prod_{\alpha =1}^n
S_{\zeta_{\alpha}}(x_{\alpha} ,t_{\alpha}) \rab
=\left( \la S_{\zeta} \rab \right)^n \exp\left\{-4\pi\sum_{\alpha,\beta,
\alpha<\beta}
\zeta_{\alpha}\zeta_{\beta}\overline{G}_{\beta, m}(x_\alpha -x_\beta,
t_\alpha -t_\beta)
\right\}~, 
\label{ancs}
\eeqn
where $\la S_{\zeta}\rab$ is given in Eq.(\ref{chcond}) and can be
rewritten as
\beqn
\la S_{\zeta}\rab= -\frac{1}{L}\exp\left \{-2\pi 
\overline{G}_{\beta,m}^{(0)}(0)
+\sum_{k>0}\frac{1}{k} -2\pi \tilde{G}_{\beta,m}(0,0)
\right\}~~. 
\label{acs}
\eeqn
Functions $\overline{G}_{\beta, m}(x,t),\overline{G}_{\beta,m}^{(0)}(t)$ and
$\tilde{G}_{\beta,m}(x,t)$ are defined in Eqs.(\ref{tevmsf}) -
(\ref{proptil}). Namely we will prove that
\beqn
&&\la \prod_{\alpha =1}^n
S_{\zeta_{\alpha}}(x_{\alpha} ,t_{\alpha}) \rabv \cr
&&=\frac{(-1)^n}{L^n}\exp \left\{-2\pi n \overline{G}_{\beta,m}^{(0)}(0)
-4\pi\sum_{\alpha,\beta, \alpha<\beta}
\zeta_{\alpha}\zeta_{\beta}\overline{G}^{(0)}_{\beta, m}
(t_\alpha -t_\beta)\right\}
\label{ancsvac1}
\eeqn
and
\beqn
&&\la \prod_{\alpha =1}^n
S_{\zeta_{\alpha}}(x_{\alpha} ,t_{\alpha}) \rabe 
=\exp\left\{ n\left(\sum_{k>0}\frac{1}{k} -2\pi \tilde{G}_{\beta,m}(0,0)
\right) \right. \cr
&&\left.-4\pi\sum_{\alpha,\beta, \alpha<\beta}
\zeta_{\alpha}\zeta_{\beta}\tilde{G}_{\beta, m}(x_\alpha - x_\beta,t_\alpha
-t_\beta)\right\}~~.
\label{ancsexc1}
\eeqn
Then due to the general statement expressed by Eq.(\ref{averbeta}) and
using the explicit forms of the Green functions given in
Eqs.(\ref{tevmsf})- (\ref{proptil}) the formula (\ref{ancs})will be
obtained.
\par
In order to prove (\ref{ancsvac1}) let us first find the representation of
the chiral operator $S_{\zeta}(x)$ in the vacuum Hilbert space ${\cal
H}_{vac}$ which has the basis $\{|E_n\ra \}$. From Eqs.(\ref{En}) and
(\ref{bosonf}) we have for the matrix element of the chiral operator:
\beqn
\bEp|S_{\zeta}(x)|\kE =
-\frac{1}{L}\sum_{N,N'}\int_0^1 d {\bar A}\int_0^1 d {\bar A'}
\Psi^{*}_{N',n'}({\bar A}')\Psi_{N,n}({\bar A})\la N', {\bar A}'|
U_{\zeta}^{\dag}
U_{-\zeta}|\kna 
\label{mecs1}
\nonumber
\eeqn
Since
\beqn
\la N', {\bar A}'|U_{\zeta}^{\dag} U_{-\zeta}|\kna = \delta_{N',N+\zeta}\delta({\bar
A}'-{\bar A})
\label{vacuu}
\eeqn
we obtain
\beqn
\bEp|S_{\zeta}(x)|\kE =
-\frac{1}{L}\sum_N \int_0^1 d {\bar A}
\Psi^{*}_{N+\zeta,n'}({\bar A})\Psi_{N,n}({\bar A}) \nonumber\\
=-\frac{1}{L}\sum_N \int_0^1 d {\bar A}
\psi^{*}_{n'}({\bar A}-N-\zeta+1/2)\psi_{n}({\bar A}-N+1/2)~,
\label{mecs2}
\eeqn
where we used Eqs.(\ref{Psi}) and (\ref{Psiho}).
Now we can again extend the integration to the whole interval $(-\infty,
\infty)$ getting (see Eq.(\ref{matnn'}))
\beqn
\bEp|S_{\zeta}(x)|\kE =
-\frac{1}{L}\int_{-\infty}^{\infty} d {\bar A}
\psi^{*}_{n'}({\bar A}-\zeta)\psi_{n}({\bar A})
=-\frac{1}{L}\bEp| e^{-i\zeta P_{vac}}|\kE ~,
\label{mecs3}
\eeqn
where
\beqn
P_{vac}= i\sqrt{\frac{\omega}{2}}(a^{\dag}-a)
\label{momvac}
\eeqn
is a momentum which corresponds to the vacuum Hamiltonian $H_{vac}$ given
in Eq.(\ref{hvac2}). So we see that the chiral operator $S_{\zeta}(x)$
in the vacuum space ${\cal H}_{vac}$ can be represented by the operator
$-\frac{1}{L}e^{-i\zeta P_{vac}}$. Using this fact we can rederive the
formula (\ref{chcvac}) straightforwardly
\beqn
\la S_{\zeta} \rabv = -\frac{1}{L}\la e^{-i\zeta P_{vac}}\rabv =
-\frac{1}{L}\la e^{\zeta \sqrt{\frac{2\pi}{Lm}}(a^{\dag}-a)} \rabv
=-\frac{1}{L}e^{-\frac{\pi}{mL}\coth\left(\frac{\beta m}{2}\right)}~~,
\eeqn
where again Eq.(\ref{tho2}) was used. Then
\beqn
&&\la \prod_{\alpha =1}^n
S_{\zeta_{\alpha}}(x_{\alpha} ,t_{\alpha}) \rabv
=\frac{(-1)^n}{L^n}\la  \prod_{\alpha =1}^n e^{iH_{vac}t_{\alpha}}
e^{-i\zeta_{\alpha}P_{vac}}e^{-iH_{vac}t_{\alpha}} \rabv \nonumber \\
&&=\frac{(-1)^n}{L^n}\la  \prod_{\alpha =1}^n
e^{\zeta_{\alpha}\left[a^{\dag}f(t_{\alpha})
- a f^{*}(t_{\alpha})\right]} \rabv ~, 
\label{ancsvac2}
\eeqn
where $f(t_{\alpha}) =\frac{2\pi}{Lm} e^{imt_{\alpha}}$ and we used
explicit form Eq.(\ref{momvac}) of the momentum  $P_{vac}$ and formula
(\ref{tho3}).
\par
Furthermore with the help of the formula : $e^Ae^B=e^{A+B+\onehalf[A,B]}$
if $[A,B]$ is a c-number we get
\beqn
\la \prod_{\alpha =1}^n
S_{\zeta_{\alpha}}(x_{\alpha} ,t_{\alpha}) \rabv
&=&\frac{(-1)^n}{L^n}
\la e^{a^{\dag}\sum_{\alpha =1}^n \zeta_{\alpha}f(t_{\alpha})
-a \sum_{\alpha =1}^n \zeta_{\alpha}f^*(t_{\alpha})}\rabv \nonumber\\
&\times& \exp\sum_{\alpha, \beta, \alpha <\beta }
\zeta_{\alpha}\zeta_{\beta}\frac{2\pi}{Lm}\sinh[im(t_{\alpha}-t_{\beta})]
~. 
\label{ancsvac3}
\eeqn
Now we can use Eq.(\ref{tho2}) with the result:
\beqn
&&\la \prod_{\alpha =1}^n
S_{\zeta_{\alpha}}(x_{\alpha} ,t_{\alpha}) \rabv \nonumber \\
&&=\frac{(-1)^n}{L^n}\exp \left \{ -\frac{\pi n}{Lm} \coth \frac{\beta
m}{2}
-\frac{2\pi}{Lm}\sum_{\alpha, \beta,\alpha < \beta}
\zeta_{\alpha}\zeta_{\beta}
\coth\frac{\beta m}{2}\cosh[im(t_{\alpha}-t_{\beta})]\right \} \nonumber\\
&&\times \exp\sum_{\alpha, \beta,\alpha < \beta}
\zeta_{\alpha}\zeta_{\beta}\frac{2\pi}{Lm}\sinh[im(t_{\alpha}-t_{\beta})]
~. 
\label{ancsvac4}
\eeqn
This is the formula (\ref{ancsvac1}) if we use Eq.(\ref{prop0}).
\par
For the excited sector from Eq.(\ref{bosonf}) we get
\beqn
&&\la \prod_{\alpha =1}^n
S_{\zeta_{\alpha}}(x_{\alpha} ,t_{\alpha}) \rabe =\exp
\left [n\sum_{k>0}\frac{1}{k} \right]
\nonumber\\
&&\times\la \prod_{\alpha =1}^n e^{ \zeta_{\alpha}\sum_{k_{\alpha}>0}\left
[\beta_{\alpha}(k_{\alpha})A^{\dag}(k_{\alpha})-\beta^*_{\alpha}
(k_{\alpha})
B^{\dag}(k_{\alpha})
-\beta^*_{\alpha}(k_{\alpha})A(k_{\alpha})+\beta_{\alpha}(k_{\alpha})
B(k_{\alpha})\right]} \rabe ~~, 
\label{ancsexc2}
\eeqn
where
\beqn
\beta_{\alpha}(k)=\frac{1}{\sqrt{k}}\left(\cosh \alpha(k) - \sinh
\alpha(k)\right)
e^{-2\pi ik\frac{x_{\alpha}}{L}+iE(k)t_{\alpha}}~.
\label{bak}
\eeqn
Now using Eq.(\ref{tho4}) and the fact that from Eq.(\ref{alpha})
\beqn
(\cosh \alpha(k) -\sinh \alpha(k))^2 =\frac{2\pi k}{LE(k)}
\label{cs2}
\eeqn
we get
\beqn
&&\la \prod_{\alpha =1}^n
S_{\zeta_{\alpha}}(x_{\alpha} ,t_{\alpha}) \rabe =
\exp\left\{n\sum_{k>0} \left(\frac{1}{k}
-\frac{2\pi}{LE(k)}\coth\frac{\beta E(k)}{2} \right) \right\} \nonumber\\
&&\times \exp \left\{ -4\pi\sum_{\alpha, \beta,\alpha < \beta}
\zeta_{\alpha}\zeta_{\beta} \sum_{k>0}\frac{\cos\left( 2\pi
k \frac{(x_{\alpha}- x_{\beta})}{L}\right )}{LE(k)} \right.\nonumber \\
&&\times \left.\left[ \coth \frac{\beta E(k)}{2}\cosh[i(t_{\alpha}-
t_{\beta})E(k)]-\sinh[i(t_{\alpha}- t_{\beta})E(k)] \right] \right\}
\label{ancsexc3}
\eeqn
and this is the formula (\ref{ancsexc1}) if we use Eq.(\ref{proptil}).
%%%%%%%%%%%%%%%%%%%%%%%%%%%%%%%%%%%%%%%%%%%%%%%%%%%%%%%%%%%%%%%%%%%%%%%%%%%
\section{Relation between path integral and Hamiltonian approaches}
In Eq.(\ref{tevmsf}) the t.e.v. of the product of two free quantum neutral
massive scalar fields $A(x,t)$ on a circle at different space-time points
was presented. In order to make a transition  to Eucledian space-time
(with the space-time coordinates $(x_1,x_2)$) we should make the following
substitutions (analytical continuation in time)
\beqn
x\rightarrow x_1,~~L\rightarrow L_1,~~t\rightarrow -ix_2,~~ \beta \rightarrow
L_2 ~~.
\label{substi1}
\eeqn
Then the propagator of this field on the Eucledian torus will be defined as
\beqn
\la A(x_1,x_2)A(x_1',x_2') \ra &=& \theta(x_2-x_2') \la A(x_1,t)A(x_1',t')\rab
|_{t=-ix_2, ~t'= -ix_2'} \cr
&+& \theta(x_2'-x_2)\la A(x_1',t')A(x_1,t) \rab |_{t=-ix_2, ~t=-ix_2'} \cr
&=&\theta(x_2-x_2') \overline{G}_{\beta,m}(x_1-x_1', -i(x_2-x_2')) \cr
&+& \theta(x_2'-x_2) \overline{G}_{\beta,m}(x_1'-x_1, -i(x_2'-x_2)) \cr
&=&\frac{1}{2 L_1}\sum_k \frac{e^{2\pi ik \frac{(x_1-x_1')}{L_1}}
\cosh \left[ E(k)\left(\frac{L_2}{2} -|x_2-x_2'|\right)\right] }
{E(k)\sinh \left(\frac{L_2E(k)}{2} \right)} \cr
&=&  \overline{G}_m(x_1-x_1', x_2-x_2')~~,
\eeqn
where we used Eqs.(\ref{tevmsf}) and (\ref{mprop}). Now if in addition
to substitutions (\ref{substi1}) we make substitutions
\beqn
\partial_x \rightarrow \partial_1, ~~ \partial_t \rightarrow i\partial_2, ~~~
j^0(x,t) \rightarrow -ij_2(x_1,x_2), ~~ j^1(x,t)\rightarrow j_1(x_1,x_2),
\nonumber \\
F_{01}(x,t)\rightarrow iF_{12}(x_1,x_2),~~~~
\gbm \rightarrow \overline{G}_m(\bar{x}_1,\bar{x}_2)
\label{substi2}
\eeqn
we will get the results obtained in the results of path integral 
Lagrangian approach in Euclidean space-time \cite{SW}, \cite {Azakov},
\cite{Azakov2} from the results obtained in this section.\\ \\

{\bf Acknowledgments}\\\\
We would like to thank H.Joos for fruitful collaboration during which 
many results of the present paper were obtained and the Abdus Salam 
International Centre for Theoretical Physics where this work was completed 
for a kind hospitality and stimulating atmosphere.\\\\

%%%%%%%%%%%%%%%%%%%%%%%%%%  my definitions  %%%%%%%%%%%%%%%%%%%%%%%%%%%

\def\mybig{\displaystyle \strut }
\def\mbig{\displaystyle }

\def\la{\langle}
\def\ra{\rangle}
\def\next{{~~~,~~~}}
\def\onehalf{ \hbox{${1\over 2}$} }
\def\jp{\hat{j}_+(k)}
\def\jpd{\hat{j}_+^{\dagger}(k)}
\def\jm{\hat{j}_-(k)}
\def\jmd{\hat{j}_-^{\dagger}(k)}
\def\pr{\hat{\psi}_R(x)}
\def\prd{\hat{\psi}_R^{\dagger}(x)}
\def\pl{\hat{\psi}_L(x)}
\def\pld{\hat{\psi}_L^{\dagger}(x)}
\def\op{\hat{O}_+(x)}
\def\om{\hat{O}_-(x)}
\def\ba{\bar{A}}
\def\hca{\hat{\cal A}}
\def\hcb{\hat{\cal B}}
\def\hpp{\hat{\varphi}_+}
\def\hpm{\hat{\varphi}_-}
\def\had{\hat{a}^\dag}
\def\ha{\hat{a}}
\def\hbd{\hat{b}^\dag}
\def\hb{\hat{b}}
%%%%%%%%%%%%%%%%%%%%%%%%%%%%%%%%%%%%%%%%%%%%%%%%%%%%%%%%%%%%%%%%%%%%%%%%

%%%%%%%%%%%%%%%%%%%%%%%%%%  my definitions  %%%%%%%%%%%%%%%%%%%%%%%%%%%

\def\la{\langle}
\def\ra{\rangle}
\def\next{{~~~,~~~}}
\def\ad{a^{\dag}}
\def\onehalf{\hbox{${1\over 2}$}}
%%%%%%%%%%%%%%%%%%%%%%%%%%%%%%%%%%%%%%%%%%%%%%%%%%%%%%%%%%%%%%%%%%%%%%%%

{\bf Appendix A. Bosonization of the fermionic operators on the circle}
\vspace*{3mm}
\setcounter{equation}{0}
\renewcommand{\theequation}{A.\arabic{equation}}

Let us first consider the bosonization of fermion field with {\it
positive chirality (right movers)} described by operators $\ha(m)$.
From Eqs.(\ref{CR}) and ({\ref{currk+-}) we have the
following commutation relations for $k>0$
\begin{subequations}
\beeq
[\jp ,\ha(m)]=-\ha(m+k),
\eneq
\beeq
[\jpd ,\ha(m)]= -\ha(m-k).
\eneq
\end{subequations}

Then for the fermion field with positive chirality
$\pr = \sum_{m}\ha(m)\phi_{R,m}(x)$ we get
\begin{subequations}
\beeq
[\jp, \pr] = -e^{-2\pi ik \frac{x}{L}}\pr,
\eneq
\beeq
[\jpd,\pr] = -e^{2\pi ik \frac{x}{L}} \pr.
\eneq
\end{subequations}
Thus $\pr$ field can be represented in the following form
\beeq
\pr=\op e^{-\hca^{\dagger}(x)}e^{\hca (x)},
\label{bos1}
\eneq
where
\beeq
\hca(x)= \sum_{k=1}^{\infty}\frac{1}{k}\jp e^{2\pi ik\frac{x}{L}}
\label{caligA}
\eneq
and $\op$ commutes with all $\jp$ and $\jpd  (k>0)$.
As $\pr$ reduces the number of right movers by one the operator $\op$ must
also have this property. So acting on the RDSS one gets:
\beeq
\op |N_+, N_-; \ba \ra =C_+(x,N_+)|N_+-1,N_-; \ba \ra, 
\label{constC_+}
\eneq
where $C_+(x,N_+)$ is a c-number, which can be found from the value of the
matrix element
\beeq
\la N_+-1,N_-;\ba | \pr |N_+, N_-; \ba'\ra.
\eneq
Since
\beeq
\la N_+-1,N_-;\ba | \ha (k) |N_+, N_-; \ba'\ra = \delta_{k,N_+-1}\delta
(\ba
-\ba') 
\label{mat1}
\eneq
we get from (\ref{bos1}), (\ref{constC_+}) and (\ref{mat1}) that
\beeq
C_+(x,N_+) = \phi_{N_+ - 1}(x)
\eneq
From (\ref{constC_+}) it follows that the operator
\beqn
\hat{U}_+ &\equiv &\op C_+^{-1}(x,N_+) \\ \nonumber
&=& \sqrt{L}\op e^{-2\pi i \frac{x}{L}(N_+-1)-ie\int_0^xA(x')dx'
+2\pi i\ba \frac{x}{L}}
\eeqn
is independent of $x$ and has the property
\beeq
\hat{U}_+\had (k)\hat{U}_+^{-1}=\had (k-1).
\eneq
We see that this 'vacuum changing operator' $\hat{U}_+$ is unitary
\beeq
\hat{U}_+\hat{U}_+^{\dag}=\hat{U}_+^{\dag}\hat{U}_+= 1
\eneq
and for an arbitrary function $f$ of the 'number operator" $f(\hat{N}_+)$
we have
\beeq
\hat{U}_+f(\hat{N}_+)= f(\hat{N}_++1)\hat{U}_+. \label{operU_+}
\eneq
Using the relation $Q_{+,v}^{reg}=N_+-\ba-1/2$ we can introduce the
(regularized) charge operator $\hat{Q}_+$ and
\beeq
\op = \frac{1}{\sqrt{L}}\hat{U}_+e^{2\pi i \frac{x}{L}\hat{Q}_+
-i\pi \frac{x}{L} + ie\int_0^xA(x')dx'}. 
\label {O_+1}
\eneq
From (\ref{operU_+}) it follows that
\beeq
[\hat{U}_+,\hat{Q}_+]=\hat{U}_+.
\eneq
Now we may formally introduce the operator $\hat{P}_+$ canonically
conjugate to $\hat{Q}_+$
\beeq
[\hat{Q}_+,\hat{P}_+] =i.
\eneq
Then $\hat{U}_+$ may be represented as
\beeq
\hat{U}_+= e^{i\hat{P}_+}
\eneq
and
\beeq
e^{i\hat{P}_+} |N_+,N_-; \ba \ra = |N_+-1,N_-; \ba \ra
\eneq
From (\ref{bos1}),(\ref{caligA}) and (\ref{O_+1}) we get the bosonization
formula
\beqn
\pr& = & \frac{1}{\sqrt{L}} \hat{U}_+e^{2\pi i \frac{x}{L}\hat{Q}_+
-i\pi \frac{x}{L}+ie \int_0^xA(x')dx'}\\ \nonumber
&\times& e^{-\sum_{k=1}^{\infty}\frac{1}{k} \jpd e^{-2\pi ik\frac{x}{L}}}
e^{\sum_{k=1}^{\infty} \frac{1}{k}\jp e^{2\pi ik\frac{x}{L}}}
\eeqn
or
\beeq
\pr = \frac{1}{\sqrt{L}} \hat{U}_+:e^{-i2\sqrt{\pi}\hpp(x)}:, \label{bos2}
\eneq
where
\beeq
\hpp(x) =\hat{\tilde \varphi}_+(x) -
\frac{1}{2\sqrt{\pi}}\left[2\pi\frac{x}{L}\hat{Q}_+ -\pi \frac{x}{L}+
e\int_0^x A(x')dx'\right] \label{phip}
\eneq
and
\beeq
\hat{\tilde \varphi}_+(x) = \frac{1}{2 \sqrt{\pi}} \sum_{k \neq 0}
\frac{1}{ik}\jpd e^{-2\pi ik \frac{x}{L}}
\label{phiplus}
\eneq
The normal ordering $: :$ is taken with respect to the currents $\jp$ and
$\jpd$, which obey the commutation relation Eq.(\ref{cra}).
The periodicity of $\pr $ given in (\ref{bos2}) follows from the equality
(see (\ref{phiplus})
\beeq
-i2\sqrt{\pi}\hpp(L) = -i2\sqrt{\pi}\hpp(0) +2\pi i(\hat{Q}_+ -1/2 +\ba)
\eneq
and the fact that acting on a vacuum state
\beeq
(\hat{Q}_+ -1/2+\ba)|N,\ba \ra = (N-1) |N; \ba \ra.
\eneq
For the fermion field with {\it negative chirality (left movers)}
described by the operators $\hb(m)$ the calculations are very similar.
Since for $k>0$
\begin{subequations}
\beeq
[\jm, \hb(m) ]= -\hb(m+k),
\eneq
\beeq
[\jmd,\hb(m)]= -\hb(m-k),
\eneq
\end{subequations}
for $\pl = \sum_m \hb(m)\phi_{L,m}(x)$ we get
\begin{subequations}
\beeq
[\jm, \pl] = -e^{-2\pi ik \frac{x}{L}}\pl
\eneq
\beeq
[\jmd, \pl] = -e^{2\pi ik \frac{x}{L}}\pl
\eneq
\end{subequations}
and $\pl $ field can be represented as
\beeq
\pl = \om e^{- \hcb^{\dag}(x)}e^{\hcb (x)},
\eneq
where
\beeq
\hcb(x) = \sum_{k=1}^{\infty}\frac{1}{k} \jmd e^{-2\pi ik \frac{x}{L}}
\label{caligB}
\eneq
$\om $ commutes with all $\jm$ and $\jmd$ and has the property
\beeq
\om |N_+,N_-; \ba \ra = C_-(x,N_-)|N_+,N_-; \ba \ra, 
\label{constC_-}
\eneq
where $C_-(x,N_-)$ is a c-number which can be found from
\beeq
\la N_+,N_- +1; \ba|\pl |N_+,N_-; \ba' \ra . \label{mat2}
\eneq
Since
\beeq
\la N_+,N_- +1; \ba| \hb(k) |N_+,N_-; \ba' \ra =
\delta_{k,N_-}\delta(\ba -\ba') \label{mat3}
\eneq
we get
\beeq
C_-(x,N_-) = \phi_{N_-}(x).
\eneq
From (\ref{constC_-}) it follows that the operator
\beqn
\hat{U}_- &\equiv & \om C_-^{-1}(x,N_-) \\ \nonumber
& =& \sqrt{L} \om e^{-2\pi i \frac{x}{L}\hat{N}_- -ie\int_0^x A(x')dx'
+ 2\pi i\ba \frac{x}{L}}
\eeqn
is independent of $x$ and has the property
\beeq
\hat{U}_-\hbd(k)\hat{U}_-^{-1} = \hbd(k+1).
\eneq
The operator $\hat{U}_-$ is a unitary operator
\beeq
\hat{U}_-\hat{U}_-^{\dag}=\hat{U}_-^{\dag}\hat{U}_- = 1
\eneq
and for an arbitrary function of the number operator $f(\hat{N}_-)$ we
have
\beeq
\hat{U}_-f(\hat{N}_-)=f(\hat{N}_- -1)\hat{U}_-.
\label{operU_-}
\eneq
Using the relation $Q_{-,v}^{reg} = -N_-+ \ba +1/2$ we write $\om$ in
terms of the (regularized) charge operator $\hat{Q}_-$
\beeq
\om = \frac{1}{\sqrt{L}}\hat{U}_-e^{-2\pi i \frac{x}{L}\hat{Q}_-
+i\pi\frac{x}{L}
+ie\int_0^xA(x')dx'}. \label{O_-1}
\eneq
From (\ref{operU_-}) it follows that
\beeq
[\hat{U}_-,\hat{Q}_-] =\hat{U}_-.
\eneq
Introducing $\hat{P}_-$ such that
\beeq
[\hat{Q}_-,\hat{P}_-] =i
\eneq
we may write $\hat{U}_-$ as follows
\beeq
\hat{U}_- = e^{i\hat{P}_-}
\eneq
and
\beeq
 e^{i\hat{P}_-}|N_+,N_-;\ba \ra = |N_+,N_-+1;\ba \ra.
 \eneq
So in analogy with (\ref{bos2}) we may write
\beeq
\pl = \frac{1}{\sqrt{L}} \hat{U}_-:e^{-i2\sqrt{\pi}\hpm(x)}:, \label{bos3}
\eneq
where
\beeq
\hpm(x) =\hat{\tilde \varphi}_-(x) +
\frac{1}{2\sqrt{\pi}}\left[2\pi\frac{x}{L}\hat{Q}_- -\pi \frac{x}{L}-
e\int_0^x A(x')dx'\right] \label{phim}
\eneq
and
\beeq
\hat{\tilde \varphi}_-(x) = \frac{1}{2 \sqrt{\pi}} \sum_{k \neq 0}
\frac{1}{ik}\jm e^{2\pi ik \frac{x}{L}} \label{phiminus}
\eneq
The normal ordering $: :$ is taken with respect to the currents $\jmd$ and
$\jm$, which obey the commutation relations Eq.(\ref{cra}).\\
The periodicity of $\pl$ given by Eq.(\ref{bos3}) follows from the
equality
\beeq
-i2\sqrt{\pi}\hpm(L)= -i2\sqrt{\pi}\hpm(0)-2\pi i(\hat{Q}_- -1/2 - \ba)
\eneq
We know that acting on a vacuum state $|N,\ba \ra$
\beeq
(\hat{Q}_- -1/2 - \ba)|N, \ba \ra = -N |N, \ba \ra.
\eneq
In order to make fields with the different chirality to anticommute we
will
introduce  for the field $\pl$ a so-called Klein factor
\beeq
\hat{C}_- =
e^{i\pi(\hat{Q}_++\hat{Q}_-)} \label{KF}
\eneq
 and $\pl$ will take a form
\beeq
\pl = \frac{1}{\sqrt{L}} \hat{C}_-\hat{U}_-:e^{-i2\sqrt{\pi}\hpm(x)}:.
\label{psilbos}
\eneq
Introduction of the Klein factor could only change the sign in the matrix
element (\ref{mat2}) or (\ref{mat3}). Such a change is permited since
different RDSS could have different phases which are not fixed a priori.\\

{\bf Appendix B. Thermodynamics of Harmonic Oscillator}

\vspace*{3mm}
\setcounter{equation}{0}
\renewcommand{\theequation}{B.\arabic{equation}}

In this appendix we present t.e.v. for  some operators in the simple
theory of a one-dimensional harmonic oscillator with  the Hamiltonian
\beeq
H=-\frac{1}{2}\frac{d^2}{dx^2}+\frac{\omega^2x^2}{2},~~~-\infty
<x<\infty~~.
\eneq
In terms of creation ($\ad$) and annihilation ($a$)
operators which obey the standard commutation relation
\beeq
[a, \ad ]=1
\eneq
it has a form
\beeq
H =\omega(\ad a+ \frac{1}{2}),
\eneq
and for the coordinate and momentum operators we have
\beeq
x=\frac{1}{\sqrt{2\omega}}(a^++a)~,
\eneq
\beeq
p=i\sqrt{\frac{\omega}{2}}(a^+-a)~,
\eneq
respectively. Eigenfunctions of the Hamiltonian in the coordinate
representation are
\beeq
\langle x|E_n\rangle = \psi_n(x) =
 \left(\frac{\omega}{\pi}\right)^{1/4}
\frac{1}{(2^n n!)^{1/2}} H_n(\sqrt{\omega}x)
e^{-\frac{\omega}{2}x^2}~.
\label{Psiho}
\eneq
Note that in this theory
\beqn
\int \psi_{n'}^*(x-\zeta)\psi_n(x)dx =\int \la E_{n'}|x-\zeta \ra
\la x|E_n\ra dx \nonumber \\
= \int\la E_{n'}|e^{-i\zeta p}|x\ra \la x|E_n \ra dx
=\la E_{n'}|e^{-i\zeta p}|E_n \ra \label{matnn'}
\eeqn
since $ e^{-i\zeta p}|x\ra = |x -\zeta \ra $.

T.e.v. of any operator $f(\ad, a)$
\beeq
\la f(\ad, a)\ra_{\beta} = Z^{-1} {\rm Tr} \Big (f(\ad,a)e^{-\beta H} \Big
)~,
\label{tho1}
\eneq
 where $Z={\rm Tr}\left( e^{-\beta H}\right)$ is a partition function, can
be easily calculated with the help of the formulae
\beeq
\la (\ad)^k a^m \ra_{\beta} = k!( \la \ad a \ra_{\beta})^k
\delta_{km}=k!(n(\omega))^k\delta_{km}~,
\eneq
\beeq
\la \ad a \ra_{\beta} =n(\omega)= \frac{1}{e^{\beta\omega}-1}~,
\eneq
\beeq
\la a \ad \ra_{\beta} = \frac{1}{1-e^{-\beta\omega}}~.
\eneq
For the partition function $Z$ we have
\beeq
Z= {\rm Tr} \Big (e^{-\beta H} \Big ) = \left[2\sinh
\frac{\beta\omega}{2}\right]^{-1}~.
\eneq

With the help of (\ref{tho1}) we obtain e.g.
\beeq
\la e^{\alpha\ad}e^{\gamma a} \ra_{\beta} =
\sum_{k,m}\frac{\alpha^k\gamma^m}
{k!m!}\la (\ad)^k a^m \ra_{\beta}
=\sum_k\frac{(\alpha\gamma)^k}{k!}(\la\ad a\ra_{\beta})^k
=e^{\alpha\gamma\la
\ad a \ra_{\beta}}=e^{\alpha\gamma n(\omega)}~.
\eneq
Then
\beqn
\la e^{\alpha\ad+\gamma a} \ra_{\beta} = \la e^{\alpha\ad}e^{\gamma a}
\ra_{\beta}
e^{\onehalf \alpha\gamma}=e^{\alpha\gamma(n(\omega)+\onehalf)}=
e^{\onehalf \alpha\gamma \coth(\onehalf \beta\omega)}~, \label{tho2}
\eeqn
where we have used BH formula: if $ [ A,B ]$ is a c-number
\beeq
e^{A+B}=e^{A}e^{B}e^{-\onehalf [A,B]}~~.
\eneq
The following formula
\beeq
e^{z\ad a}f(\ad,a)e^{-z\ad a}= f(e^z\ad,e^{-z} a)~. \label{tho3}
\eneq
can be used for finding time dependence of the operators and in
the calculations of thermodynamical correlators.

For two independent harmonic oscillators with the same frequencies, which
are described  by the Hamiltonian $ H = \omega \ad a + \omega b^{\dag} b
$
(we omit the constant terms which are not important for t.e.v.) with the
help of Eqs.(\ref{tho2})and (\ref{tho3}) we can get
\beqn
&&\la \prod_{\alpha =1}^n e^{it _{\alpha}H}
e^{\gamma_{\alpha}a^{\dag} - \gamma_{\alpha}^{\ast}a
-\gamma_{\alpha}^{\ast}b^{\dag} + \gamma_{\alpha}b}
e^{-it _{\alpha}H} \ra_{\beta}\cr
&&=\exp \Big \{ - \sum_{\alpha = 1}^{n}|\gamma_{\alpha}|^2 \coth
\frac{\beta \omega}{2} 
- \sum_{\alpha, \delta (\alpha < \delta)}
(\gamma_{\alpha}\gamma_{\delta}^{\ast}+\gamma_{\alpha}^{\ast}
\gamma_{\delta})\Big[\coth \frac{\beta \omega}{2} \cosh \omega
(it_{\alpha}-it_{\delta}) \cr 
&&- \sinh \omega (it_{\alpha}-it_{\delta})
\Big] \Big \} 
\label{tho4}
\eeqn
The second sum in $\{\ldots \}$ exists only if $n \geq 2$.

{\bf Appendix C}\\
\setcounter{equation}{0}
\renewcommand{\theequation}{C.\arabic{equation}}

 We have two sums
\beqn
S_1=\sum_{n=0}^{\infty}e^{-(\beta-it)nm}\Psi_{N+1,n}^{\ast}(\bar{A})
\Psi_{N_1,n}(\bar{A}_1)
\eeqn
and
\beqn
S_2=\sum_{n=0}^{\infty}e^{-it nm}\Psi_{N_1+1,n}^{\ast}(\bar{A}_1)
\Psi_{N,n}(\bar{A})
\eeqn
Let us first calculate $S_2$. From (\ref{Psi})) we have
\beqn
S_2=\left (\frac{\omega}{\pi} \right)^{1/2} e^{-\frac{x^2}{2}-\frac{y^2}{2}}
\sum_{n=0}^{\infty}H_n(x)H_n(y)\frac{\xi^n}{2^n n!},
\eeqn
where
\[ x\equiv \sqrt{\omega}(N_1-\bar{A}_1+1/2),~~~
y\equiv \sqrt{\omega}(N-\bar{A}-1/2),~~~
\xi\equiv e^{-itm}     \]
Using the Mehler's formula (\ref{Meh}) we get
\beqn
S_2=\left (\frac{\omega}{\pi} \right)^{1/2}(1-\xi^2)^{-1/2}\exp[2xy\delta
- (x^2+y^2)\gamma ]~~, \label{S2}
\eeqn
where
\[\gamma \equiv \frac{1+\xi^2}{2(1-\xi ^2)}=\frac{1}{2}\coth (it
m),~~~~\delta \equiv \frac{\xi}{1-\xi ^2}=\frac{1}{2\sinh(itm)} \]
For the first sum we get the similar expression
\beqn
S_1=\left (\frac{\omega}{\pi}
\right)^{1/2}(1-\tilde{\xi}^2)^{-1/2}\exp[2\tx
\ty\td - ({\tx}^2+{\ty}^2)\tg ]~~, \label{S1}
\eeqn
where
\[ \tx=\sqrt{\omega}(N-\bar{A}+1/2),~~ ~~
\ty=\sqrt{\omega}(N_1-\bar{A}_1-1/2),
~~~\tilde{\xi}=e^{-(\beta-it)m},\]
\[ \tg \equiv \frac{1+\tilde{\xi}^2}{2(1-\tilde{\xi}^2)},~~~
\td \equiv \frac{\tilde{\xi}}{1-\tilde{\xi}^2} \] \\

{\large \bf Integrations}

Let us first consider the integral with respect to $\bar{A}$ and  the sum
with respect to $N$. The variables which contain $\bar{A}$ and $N$ are
$\tx$ and $y$ $(\tx=y+\sqrt{\omega})$. From (\ref{(1)}), (\ref{S1}) and
(\ref{S2}) it follows
that we should calculate the sum and the integral

\[I_1\equiv\sum_{N}\int_0^1 d\bar{A}e^{2xy\delta-y^2\gamma+
2\tx \ty \td -{\tx}^2 \tg} \]

\[=e^{2\sqrt{\omega} \ty \td -\omega \tg}
\sum_N\int_0^1 d\bar{A} e^{-y^2(\gamma+\tg)+2y(x\delta+\ty \td -
\sqrt{\omega}\tg )} \]
Using Manton's perodicity condition (\ref{pc}) we can again substitute
\[ \sum_N \int_0^1d\bar{A} \rightarrow \int_{-\infty}^{\infty}d \bar{A} \]
and
\beqn
I_1=\sqrt{\frac{\pi}{\omega(\gamma+\tg)}}e^{2\sqrt{\omega}\ty \td -
\omega \tg }e^{\frac{(x\delta +\ty \td -\sqrt{\omega}\tg
)^2}{\gamma+\tg}}.
\eeqn
Now we should do the summation with respect to $N_1$ and integration with
respect to ${\bar A}_1$. The variables which contain $N_1$ and ${\bar A}_1$
are $\ty$ and $x$ $(x=\ty +\sqrt{\omega})$ and we must calculate
\[ I=\sum_{N_1}\int_0^1 d\bar{A}_1 I_1(x, \ty)e^{-x^2\gamma-{\ty}^2 \tg} \]
The result is
\beqn
I=\sqrt{\frac{\pi}{\omega(\gamma+\tg)}} \sqrt{\frac{\pi}{a}}
e^{-\omega(\gamma+\tg)}e^{\frac{b^2}{a}}e^{\frac{\omega (\delta - \tg)^2}
{\gamma+\tg}}~,
\eeqn
where
\beqn
a\equiv\frac{\omega \left[(\gamma +\tg)^2-(\delta +\td )^2 \right]}
{\gamma+\tg}~~, \label{a}
\eeqn
\beqn
b\equiv\frac{\omega [(\delta +\td)(\delta -\tg)-(\gamma -
\td)(\gamma+\tg)]}
{\gamma+\tg} ~~. \label{b}
\eeqn
So for $\la(1)\ra_{vac}$ in (\ref{(1)}) we finally get
\beqn
\la(1)\ra_{vac}= Z_{vac}^{-1}\frac{1}{L^2}\frac{1}{\sqrt{(\gamma+\tg)^2-
(\delta +\td)^2}}\frac{1}{\sqrt{(1-\xi^2)(1-\tilde{\xi}^2)}} \cr
\times \exp \left [ -\omega(\gamma+\tg)+\omega \frac{(\delta -\tg)^2}
{\gamma +\tg} +\frac{b^2}{a} \right] ~~. \label{(1)1}
\eeqn
The rest of the calculations is just to do some simplfications dealing
with hyperbolic functions
\beqn
\delta+ \td =A_1(\gamma+\tg),~~~\delta -\tg =A_2(\gamma+\tg),~~~
\gamma - \td =A_3(\gamma+\tg),
\eeqn
where
\beqn
A_1\equiv\frac{\cosh\left(\frac{\beta}{2}-it \right)m }{\cosh
\frac{\beta}{2}m}
\eeqn
\beqn
A_2\equiv\frac{\sinh(\beta-it)m -\cosh(\beta-it)m\sinh itm}
{\sinh\beta m}
\eeqn
\beqn
A_3\equiv\frac{\cosh itm \sinh(\beta-it)m -\sinh itm}
{\sinh\beta m}
\eeqn
Note that
\beqn
A_3=A_2-A_1+1~~. \label{A3}
\eeqn
The expression for $a$ in (\ref{a}) is simple
\beqn
a=\omega \tanh\frac{\beta m}{2},
\eeqn
\beqn
\gamma+\tg =\frac{a}{\omega(1-A_1^2)}=\frac{\sinh\frac{\beta m}{2}
\cosh\frac{\beta m}{2}}{\sinh(\beta -it)m \sinh itm}
\eeqn
For $b$ defined in (\ref{b}) we have
\[b= \omega [A_1A_2-A_3](\gamma+\tg) \]
and using (\ref{A3})
\beqn
b=\omega (A_1-1)(1+A_2)(\gamma+\tg)
\eeqn
and for $b^2/a$ we get
\beqn
\frac{b^2}{a}=\omega \left(\frac{1-A_1}{1+A_1}\right)(1+A_2)^2
(\gamma+\tg)
\eeqn
and for the whole exponent in (\ref{(1)1}) we get
\beqn
-\omega (\gamma+\tg)+\omega \frac{(\delta -\tg)^2}{\gamma+\tg}
+\frac{b^2}{a}=2a\frac{(1+A_2)}{(1+A_1)}\frac{(A_2-A_1)}{(1-A_1^2)}\cr
=-\frac{\omega}{2}\coth\frac{m\beta}{2}-\frac{\omega}{2}
\left (\coth\frac{m\beta}{2}\cosh m it -
\sinh m it \right)
\eeqn
and this is the exponent in (\ref{nvac}) when $\zeta = -1$.\\
For the factor in front of the exponential function in (\ref{(1)1}) we
should
use
\beqn
(\gamma+\tg)^2 -(\delta +\td)^2=
\frac{\sinh^2\frac{\beta m}{2}}
{\sinh(\beta -it)m \sinh itm}
\eeqn
\beqn
\frac{1}{\sqrt{(1-\xi^2)(1-\tilde{\xi}^2)}}=
\frac{1}{\sqrt{\xi \tilde{\xi} \left(\frac{1}{\xi}-\xi
\right) \left(\frac{1}{\tilde{\xi}}-\tilde{\xi} \right)}}\cr
=\frac{e^{\frac{\beta m}{2}}}{\sqrt{4\sinh(\beta -it)m \sinh itm}}
 \eeqn
The vacuum partition function in these calculations is
\beqn
Z_{vac}=\frac{e^{\frac{\beta m}{2}}}{2\sinh\frac{\beta m}{2}}
\eeqn
since in the  Hamiltonian $H_{vac}$ we omitted the constant terms.

\end{document}